\documentclass{aa}

\usepackage{xcolor}
\usepackage{multirow}
\usepackage{txfonts}
\usepackage{graphicx}
\usepackage{fontawesome}
\usepackage{placeins}
\usepackage{hyperref}

\newcommand{\sn}{SNe Ia}
\newcommand{\mpc}{h^{-1}\text{$\cdot$Mpc}}
\newcommand{\git}[2]{\href{https://github.com/#1}{\faGithub}\footnote{\label{#1}#2\url{https://github.com/#1}}}
\newcommand{\irow}[1]{
  \begin{matrix}(\,#1\,)\end{matrix}
}

\newcommand{\Mll}{M_{\mathrm{ab},n}^{\ell_1, \ell_2}}
\newcommand{\Bi}{F_{\mathrm{ab},n}}
\newcommand{\Bvv}{F_{\mathrm{vv}, 0}}
\newcommand{\Ai}{w_{\mathrm{ab},n}}
\newcommand{\Avv}{w_{\mathrm{vv},0}}
\newcommand{\Aai}{w_{\mathrm{a},n}}
\newcommand{\Abi}{w_{\mathrm{b},n}}
\newcommand{\Av}{w_{\mathrm{v},0}}
\newcommand{\Aii}{w_{\mathrm{ab},i}}
\newcommand{\Pabi}{\mathcal{P}_{\mathrm{ab},n}}
\newcommand{\Ml}{M_{\mathrm{ab},n}^{\ell^\prime}}
\newcommand{\Mlnp}{M_{\mathrm{ab},n}^{\ell}}
\newcommand{\Lg}[1]{L_{\ell_{#1}}}
\newcommand{\Cab}{C_\mathrm{ab}(\mathbf{r_1},\mathbf{r_2})}
\newcommand{\Cabnor}{C_\mathrm{ab}}
\newcommand{\Nlll}{N_{\mathrm{ab},\ell}^{\ell_1,\ell_2}}
\newcommand{\Cabi}{C_{\mathrm{ab},n}(\mathbf{r_1},\mathbf{r_2})}
\newcommand{\Ylm}[2]{Y_{\ell_{#1}^{#2}m_{#1}^{#2}}}
\newcommand{\Ylms}[2]{Y^*_{\ell_{#1}^{#2}m_{#1}^{#2}}}
\newcommand{\Gaunt}{G_{\ell, \ell_1, \ell_2}^{m, m_1, m_2}}
\newcommand{\intsphere}{\int_{0}^{\infty} \int_{\alpha^\prime=0}^\pi \int_{\phi^\prime=0}^{2 \pi}  k^2 dk \sin{\alpha^\prime}   d \alpha^\prime d \phi^\prime}
\newcommand{\normv}[1]{\hat{\mathbf{{#1}}}}

\begin{document}

   \title{Generalized framework for likelihood-based field-level inference of growth rate from velocity and density fields}
   \titlerunning{Generalized likelihood-based field-level inference framework}

   \author{Corentin Ravoux\inst{1,2}\fnmsep\thanks{Coresponding author: corentin.ravoux@clermont.in2p3.fr}
          \and
          Bastien Carreres\inst{2,3}
          \and
          Damiano Rosselli\inst{2}
          \and
          Julian Bautista\inst{2}
          \and
          Anthony Carr\inst{4}
          \and
          Tyann Dumerchat\inst{2}
          \and
          Alex G. Kim\inst{5}
          \and
          David Parkinson\inst{4}
          \and
          Benjamin Racine\inst{2}
          \and
          Dominique Fouchez\inst{2}
          \and
          Fabrice Feinstein\inst{2}
          }
         \institute{
         Université Clermont-Auvergne, CNRS, LPCA, 63000 Clermont-Ferrand, France
         \and Aix Marseille Université, CNRS/IN2P3, CPPM, Marseille, France
         \and Department of Physics, Duke University Durham, NC 27708, USA
         \and Korea Astronomy and Space Science Institute, 776 Daedeok-daero, Yuseong-gu, Daejeon 34055, South Korea
         \and Lawrence Berkeley National Laboratory, 1 Cyclotron Road, Berkeley, CA 94720, USA
}
  \abstract
  {Measuring the growth rate of large-scale structures ($f$) as a function of redshift has the potential to break degeneracies between modified gravity and dark energy models, when combined with expansion-rate probes. Direct estimates of peculiar velocities of galaxies have attracted interest as a means of estimating $f\sigma_8$. In particular, field-level methods can be used to fit the field nuisance parameter along with cosmological parameters simultaneously. This article aims to provide the community with a unified framework for the theoretical modeling of the likelihood-based field-level inference by performing fast field covariance calculations for velocity and density fields. Our purpose is to lay the foundations for a nonlinear extension of the likelihood-based method at the field level. We have developed a generalized framework, implemented in the dedicated software \texttt{flip} to perform a likelihood-based inference of $f\sigma_8$. We derived a new field covariance model, which includes wide-angle corrections. We also included the models previously described in the literature inside our framework. We compared their performance against ours, and we validated our model by comparing it with the two-point statistics of a recent N-body simulation. The tests we performed have allowed us to validate our software and determine the appropriate wavenumber range to integrate our covariance model and its validity in terms of separation. Our framework allows for a wider wavenumber coverage to be used in our calculations than in previous works, which is particularly interesting for nonlinear model extensions. Finally, our generalized framework allows us to efficiently perform a survey geometry-dependent Fisher forecast of the $f\sigma_8$ parameter. We show that the Fisher forecast method we developed gives an error bar that is $30$ \% closer to a full likelihood-based estimation than a standard volume Fisher forecast.}


\maketitle

\section{Introduction}

In recent years, a large number of cosmological analyses have used peculiar velocities of galaxies to infer the growth rate of large-scale structures, denoted as $f\sigma_8$ (see e.g.,~\cite{Turner2024} for a recent review). These velocities can be determined using a variety of methods, all of which involve determining redshift-independent distance indicators and spectroscopic redshifts.

Two of the most widely used distance indicators are the Tully-Fisher (TF) relation \citep{tully_1977} and the Fundamental plane (FP) relation~\citep{Djorgovski_1987}. Two large recent samples of peculiar velocities determined with such methods are the Cosmicflows project for the TF relation~\citep{kourkchi_2022} and SDSS-PV for the FP relation \citep{Howlett2022}. Type Ia supernovae (\sn) have also been considered as another distance indicator to determine the peculiar velocities~\cite{howlett_measuring_2017,huterer_testing_2017,scolnic_next_2019,kim_complementarity_2020,carreres_growth-rate_2023}.

Peculiar velocities can be combined with measurements of redshift-space distortion (RSD) of galaxies in the same volume, allowing us to reach stronger constraints on the growth rate. In particular, this can be done with compressed two-point statistics such as the density and momentum correlation function, power spectra, or the average pair-wise velocities \citep{Ferreira1998,Dupuy2019,Howlett2019,Turner2022,Qin2025}. An alternative class of methodologies infer cosmological parameters directly from the velocity and density fields, without compression. The density-velocity comparison method~\citep{Springob2014,Carrick2015,Boruah2020,Said2020,Qin2023,Boubel2023} compares the observed velocity field to a predicted model based on an observed galaxy density field, typically using reconstruction techniques. A more complex method often referred to as forward modeling, or simulation-based inference \citep{boruah_reconstructing_2021,Valade2022a,Valade2022b,Pfeifer2023}, consists of evolving initial conditions of density and velocity fields using some theoretical model, generally through perturbation theory or simplified simulations, to fit observations.

The focus of this work is on another type of commonly used method: the likelihood-based field-level estimator~\citep{Johnson2014,howlett_measuring_2017,adams_improving_2017,huterer_testing_2017,adams_joint_2020,lai_using_2022,carreres_growth-rate_2023}. It consists of calculating the theoretical correlations for every pair of positions in the considered field (density or velocities), as a function of cosmological and nuisance parameters. Those parameters are varied to maximize the likelihood of the model given the data, which is assumed to be drawn from a multivariate Gaussian distribution. This maximization can be performed in a statistical or Bayesian way. Note that in this method, correlations are not compressed into binned statistics such as the correlation function or the power spectrum, and the theoretical covariance matrix corresponds to the field covariance.

Compared to traditional two-point statistics, the likelihood-based estimator allows for simultaneous fitting of the analysis parameters used to derive the fields, along with cosmological and nuisance parameters, while catching potential degeneracies between parameters. To give a more practical example, when considering SN~Ia~peculiar velocities, the velocities are estimated from Hubble diagram residuals, which depend on standardization parameters. The parameter of interest is the growth rate of structures, but some nuisance parameters are needed to perform the fit. In addition, the likelihood-based method allows one to capture two-point correlations in the field without compression, maximizing the potential extracted cosmological information. This method also allows one to add elements of the model predictions not described by Gaussian statistics, such as non-Gaussian likelihood corrections. The likelihood-based method fits a small number of parameters compared, for instance, to forward-modeling techniques in which the field data points considered are also free parameters, increasing the computational cost and complexity of the inference. However, one disadvantage of the likelihood-based method is its expensiveness in computation time and memory, as the likelihood calculation generally requires a covariance matrix inversion, which scales as $N^2$, where $N$ is the number of objects or mesh cells. The latter can scale up to approximately $10,000$ elements compared to compressed statistics, where the data vector is typically smaller than O(100) elements. Furthermore, in previous studies, the likelihood-based estimator generally computes the field theoretical correlations with linear models, and its expansion to a nonlinear model yields a higher mathematical complexity than compressed statistics and is generally computationally expensive.

To address the aforementioned issues of the likelihood-based field-level estimator, we developed a generalized framework that aims to reproduce the past works models~\cite{adams_improving_2017,adams_joint_2020, lai_using_2022,carreres_growth-rate_2023}, with or without wide-angle modeling, and to extend them to more complex field covariance models. The mathematical foundations of this framework allows one to treat all those models in a consistent way, with adapted algorithmic optimization. This framework has been implemented in the Python package \texttt{flip} \git{corentinravoux/flip}{flip: Field Level Inference Package }, which is a generalized extension of an early version used for likelihood-based inference of \sn~peculiar velocities in~\cite{carreres_growth-rate_2023}. The purpose of this paper is to present the mathematical formalism used in this package and all the applications currently included.

This work paves the way for likelihood-based field-level method improvements, such as nonlinear power spectra models, the direct inclusion in the analysis of observational effects at the field level, or the unified treatment of several velocity fields (FP, TF, \sn) with galaxy density field estimates.

This article is organized as follows. In section \ref{sec:covariance}, we detail our generalized framework for field-level covariance computation, in a wide-angle or plane-parallel configuration. Section~\ref{sec:mle} gives an overview of the likelihood-based field-level method, along with previous and new covariance models developed in this study. Section~\ref{sec:validation} shows the different validation tests performed on the \texttt{flip} algorithm. In section~\ref{sec:fisher}, we apply our covariance calculation to create a survey-dependent Fisher forecasting tool. Finally, section~\ref{sec:prospect} discusses future improvements of the \texttt{flip} package.

\section{Generalized theoretical covariance framework}
\label{sec:covariance}

\begin{figure}
	\includegraphics[width=0.95\columnwidth]{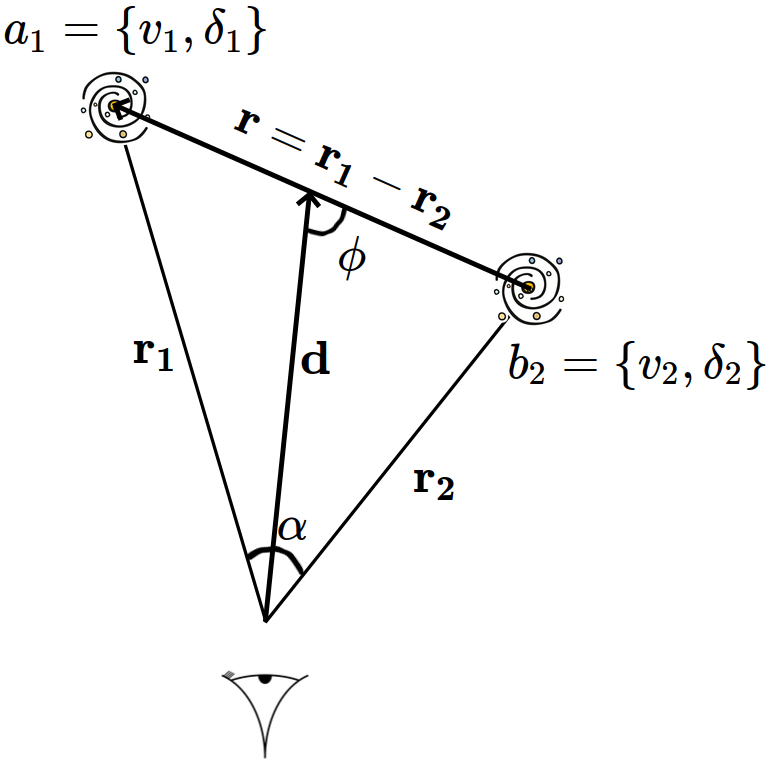}
    \caption{Schematic representation of the two field elements, $a_1$ and $b_2$, for which we want to compute the theoretical correlation. Those two fields can be the peculiar velocity of a considered galaxy or group of galaxies, or the galaxy density field itself. The definition of the vector, $\mathbf{d}$, and consequently the angle, $\phi$, depends on the chosen wide-angle definition.}
    \label{fig:angles}
\end{figure}

When performing a likelihood-based inference from two fields, denoted as a and b, which can be for example radial velocities, $v$, galaxy number density, $\delta$, velocity divergence, $\theta$, or logarithmic distance ratios, $\eta$, we assume that the fields can be described by perturbations on a Gaussian random field that depends on a theoretical covariance denoted as $\Cabnor$. In the method presented here, this covariance matrix is estimated in an analytical way by computing the theoretical correlation between each considered field. Note that the covariance terms here refer to the covariance at the field level, and is different from the covariance matrix used to fit correlation functions and power spectra. The main objective of this section is to derive a framework for fast computation of theoretical correlations of velocities and densities.

\subsection{Coordinate definition}

The coordinates used to develop our framework are shown in Fig.~\ref{fig:angles}. We consider two points of a given field (e.g., density or velocity) with distances to the observer represented by the vectors $\mathbf{r_1}$ and $\mathbf{r_2}$, respectively. Their separation vector is $\mathbf{r} = \mathbf{r_1} - \mathbf{r_2}$ and their angular separation is represented by the angle $\alpha$. We also define the $\alpha$-angle bisector $\mathbf{d}$ and the angle $\phi$ between $\mathbf{d}$ and $\mathbf{r}$. 

In this coordinate definition, we are considering the general case where the individual line-of-sight $\mathbf{r_1}$ and $\mathbf{r_2}$ are not parallel. Calculating the correlation in the non-parallel case is often referred to as wide-angle modeling. The plane-parallel approximation is when the vectors $\mathbf{r_1}$, $\mathbf{r_2}$, and $\mathbf{d}$ are all considered to be parallel. We note that in the wide-angle case, there exist several potential definitions of the $\mathbf{d}$ vector, considered as the line-of-sight reference for correlation calculation~\cite{beutler_interpreting_2019}. Changing the definition of this vector can have an impact on the modeling of correlation calculation (see~\cite{Castorina2019}). The \texttt{flip} package implements the bisector (the one used in this article), midpoint and endpoint definitions of $\mathbf{d}$:

\begin{equation}
\label{eq:d_def}
\begin{split}
    &\mathbf{d}_{\mathrm{bisector}} = \frac{r_1 r_2}{r_1 + r_2} \left(\frac{\mathbf{r_1}}{r_1} + \frac{\mathbf{r_2}}{r_2} \right) \\
    &\mathbf{d}_{\mathrm{midpoint}} = \frac{\mathbf{r_1} + \mathbf{r_2}}{2} \\
    &\mathbf{d}_{\mathrm{endpoint}} = \mathbf{r_1}\, .
\end{split}
\end{equation}

We computed the covariance of the fields a and b, respectively evaluated at positions $\mathbf{r_1}$ and $\mathbf{r_2}$. Note that this formalism is kept general on purpose and can be extended to any considered correlated field described by some power spectrum model. We used the following Fourier transform convention:

\begin{equation}
  \mathrm{a}(\mathbf{r})  =\frac{1}{(2\pi)^3} \int_\mathbf{\mathbf{k}} d^3\mathbf{k}\ \tilde{\mathrm{a}}(\mathbf{k}) e^{i\mathbf{k}\cdot\mathbf{r}}.
\end{equation}

In addition we define the individual line-of-sight angle cosine $\mu_i = \normv{k}\cdot\normv{r}_i$ where $\normv{k} = \mathbf{k} / k$ and $\normv{r}_i= \mathbf{r}_i / r_i$ with $k$ and $r_i$ are the norms of the vectors $\mathbf{k}$ and $\mathbf{r}_i$, respectively. We also define the cosine angle $\mu = \normv{k}\cdot \normv{d} = \cos(\phi)$.
The power spectrum is defined by
\begin{equation}
    \langle \tilde{\mathrm{a}}(\mathbf{k})\tilde{\mathrm{b}}^*(\mathbf{k'})\rangle = (2\pi)^3 P_\mathrm{ab}(k, \mu_1, \mu_2) \delta_D(\mathbf{k}-\mathbf{k'}),
\end{equation}

\noindent where $\delta_D$ is the Dirac delta function. The corresponding correlation is then defined as the Fourier transform of the power spectrum as:

\begin{equation}
\label{eq:cov}
\Cab = \frac{1}{(2 \pi)^3} \int_{\mathbf{k}}d^3 \mathbf{k}  P_{\mathrm{ab}}(k, \mu_1, \mu_2) e^{i \mathbf{k}\cdot\mathbf{r}} .
\end{equation}

\subsection{Wide-angle framework}

In the general wide-angle formalism, we consider that for every pair of fields a and b, we can decompose the power spectrum model as 

\begin{equation}
\label{eq:Pwflip}
P_{\mathrm{ab}}(k, \mu_1, \mu_2) = \sum_{n} \Ai \Bi(k, \mu_1, \mu_2)  \Pabi(k) \, ,
\end{equation}

\noindent where $n$ stands for the index of the term in the power spectrum model.

The logic of this decomposition is to have a linear combination of geometrical terms $\Bi(k, \mu_1, \mu_2)$ that contain the angular information to account for wide-angle effects, and isotropic power spectrum terms $\Pabi(k)$ that can be computed by a Boltzmann solver software. The coefficient of the linear decomposition $\Ai$ are the parameters we want to fit in the likelihood-based framework.

The main interest of this decomposition is to compute separately each covariance term just once and only vary $\Ai$ when maximizing the likelihood. Furthermore, integrals of the geometrical part ($\Bi(k, \mu_1, \mu_2)$) can be performed analytically while the power spectrum part ($\Pabi(k)$) can be integrated numerically in an algorithmically optimized way. Note that this decomposition might not be directly applicable for more complex nonlinear models, but this is beyond the scope of this paper.

The corresponding covariance matrix given in equation \ref{eq:cov} can then be written

\begin{equation}
    \label{eq:sum_cov}
    \Cab = \sum_n \Ai \Cabi \, ,
\end{equation}

\noindent where $\Cabi$ can be expressed as an integral over the spherical coordinates of the vector $\mathbf{k}$, defined in Eq.~\ref{eq:spherical_decomposition}, as

\begin{equation}
\label{eq:Cabi}
    \begin{split}
        \Cabi &= \frac{1}{(2 \pi)^3} \int_{\mathbf{k}}d^3 \mathbf{k}  \Bi(k, \mu_1, \mu_2)  \Pabi(k) e^{i \mathbf{k}\cdot\mathbf{r}} \\
        &= \int_{0}^{\infty} \frac{k^2 dk}{(2\pi)^3}  \Pabi(k)  \\ 
        &\times\int_\Omega d\Omega \Bi(k, \mu_1, \mu_2)e^{i\mathbf{k}\cdot\mathbf{r}}\, .
    \end{split}
\end{equation}

Using the Legendre polynomial expansion (Eq.~\ref{eq:LegendreExp}) on $\Bi (k, \mu_1, \mu_2)$ with respect to $\mu_1$ and $\mu_2$ we get
\begin{align}
\label{eq:B}
    \Bi(k, \mu_1, \mu_2) &= \sum_{\ell_1, \ell_2}\frac{(2\ell_1 + 1)(2\ell_2+1)}{4} \Lg{1}(\mu_1) \Lg{2}(\mu_2) \times\\
       &\int_{-1}^{1} \int_{-1}^{1}d\mu_1^\prime d\mu_2^\prime  \Bi(k, \mu_1^\prime, \mu_2^\prime) \Lg{1}(\mu_1^\prime) \Lg{2}(\mu_2^\prime),
\end{align}

\noindent where the $L_\ell$ are the Legendre polynomials. For conciseness, we define a wavenumber term $\Mll(k)$ that contains the analytical integration of the $\Bi (k, \mu_1, \mu_2)$ term over $\mu_1$ and $\mu_2$ such that

\begin{equation}
    \label{eq:mlldef}
       \Mll(k)=\frac{1}{4}
       \int_{-1}^{1} \int_{-1}^{1}d\mu_1^\prime d\mu_2^\prime  \Bi(k, \mu_1^\prime, \mu_2^\prime) \Lg{1}(\mu_1^\prime) \Lg{2}(\mu_2^\prime).
\end{equation}

Combining Eq.~\ref{eq:Cabi}, Eq.~\ref{eq:B} and using the plane-wave expansion (Eq.~\ref{eq:planewave}) to decompose the $e^{i\mathbf{k}\cdot\mathbf{r}}$ term, we obtain
\begin{equation}
    \label{eq:Cabil}
    \begin{split}
       \Cabi &= \int_{0}^{\infty} \frac{k^2 dk}{(2\pi)^3} \Pabi(k) \\ 
       &\times \sum_{\ell, \ell_1, \ell_2} i^\ell j_\ell(kr)(2\ell + 1)(2\ell_1+1)(2\ell_2+1)\Mll(k) \\
      &\times \int_\Omega d\Omega \Lg{}(\mathbf{\hat{k}}\cdot\mathbf{\hat{r}})\Lg{1}(\mu_1)\Lg{2}(\mu_2)\, .
    \end{split}
\end{equation}

Using Eq.~\ref{eq:intLLL} to express the three Legendre polynomial product, we obtain

\begin{equation}
    \begin{split}
       \Cabi &=\sum_{\ell, \ell_1, \ell_2} (4 \pi)^2 \sum_{m, m_1, m_2} \Gaunt \Ylms{}{}(\hat{\mathbf{r}})\Ylms{1}{}(\hat{\mathbf{r_1}})\Ylms{2}{}(\hat{\mathbf{r_2}})\\
       &\times i^\ell \int_{0}^{\infty} \frac{k^2 dk}{2\pi^2} \Pabi(k) \Mll(k)j_\ell(kr)\, .
    \end{split}
\end{equation}

For conciseness, we write the coefficient $\Nlll$ containing the linear combination of spherical harmonics such that

\begin{equation}
    \label{eq:nlldef}
    \Nlll(\normv{r},\normv{r}_1,\normv{r}_2) = (4 \pi)^2 \sum_{m, m_1, m_2} \Gaunt \Ylms{}{}(\hat{\mathbf{r}})\Ylms{1}{}(\hat{\mathbf{r_1}})\Ylms{2}{}(\hat{\mathbf{r_2}}).
\end{equation}

We can explicitly express the dependency of the spherical harmonic terms inside the $\Nlll$ terms considering the definition of angles in figure~\ref{fig:angles}:

\begin{equation}
    \begin{split}
    \Nlll(\normv{r},\normv{r}_1,\normv{r}_2) = (4 \pi)^2 &\sum_{m, m_1, m_2} \Gaunt \Ylms{}{}(\pi - \phi,0) \\
    &\times \Ylms{1}{}(\alpha /2 ,0)\Ylms{2}{}(\alpha /2 ,\pi).
    \end{split}
\end{equation}

The choice regarding the $\pi$ terms in the orientation of the angles depends entirely on the definition of $\mathbf{r}$. Furthermore, the value of the $\phi$ angle depends on the definition of the $\mathbf{d}$ vector, and can have an impact on field covariance modeling.

We define the zeroth-order Hankel transform $\mathcal{H}_\ell$ in a more cosmologically relevant form following \cite{Karamanis2021}  as 

\begin{equation}
    \mathcal{H}_\ell\left[f(k)\right](r) = i^\ell \int_{0}^\infty\frac{k^2 dk}{2\pi^2}j_\ell(kr)f(k).
\end{equation}

Finally, the individual covariance terms can be written:
    
\begin{equation}
      \label{eq:cov_flip_wa}
       \Cabi  = \sum_{\ell, \ell_1, \ell_2}\Nlll(\phi,\alpha) \mathcal{H}_\ell\left[\Pabi(k) \Mll(k)\right](r)\, .
\end{equation}

This equation is in an algorithmically optimized form for a fast computation of the field covariance matrix. The integration of the $\Nlll$ and $\Mll$ terms can be performed analytically, and a numerical integration is performed over the wavenumber $k$. The latter can easily be accelerated as it is in the form of a Hankel transform. 

In practice, for $N$ field points, the integral of Eq.~\ref{eq:cov} has to be computed $N (N+1)/ 2$ times if we include the diagonal term. This process is computationally intensive, particularly in the context of the new generation of surveys (e.g., the Legacy Survey of Space and Time Rubin-LSST~\citep{ivezic_lsst_2018} or the Dark Energy Spectroscopic Instrument DESI~\citep{desi_collaboration_desi_2016,desi_collaboration_desi_2016-1,martini_overview_2018}). The \texttt{flip} software, with the formalism developed here, proposes an efficient way to compute the covariance matrix using Hankel transforms and parallelized processes.

In our derivation, we did not consider the redshift at which the model was calculated. For small surveys concentrated on a limited redshift range, accounting for this might not be necessary, but this is not the case for the new aforementioned surveys. In particular, the parameters $\Ai$ can depend on the redshift of the two considered points. Furthermore, the power spectra, $\Pabi$, computed with Boltzmann solvers also depend on redshift. Assuming that the power spectrum redshift dependency can be factorized, we developed an option to account for different redshift between fields a and b and provide details in appendix \ref{appendix:redshift_dep}. 

We have only considered a covariance calculation between scalar fields, either between components of a vector field or a scalar field itself. Our formalism can be extended to correlations between vector fields, which means computing covariances between all components of each vector, yielding a covariance tensor. This derivation is given in appendix \ref{appendix:vector_cov}.

\subsection{Plane-parallel approximation}

In the case in which the angles between the two points of the considered field are small, we can safely use the plane-parallel approximation and simplify the modeling of the field covariance matrix. Here, we show how this approximation is implemented in \texttt{flip}. The principal purpose of this implementation is to have reference models and to compare them with previous implementations \citep{adams_improving_2017,adams_joint_2020}.

In the plane-parallel approximation it is assumed that the two galaxies are on the same plane at the same distance, $d$. Then we have $\mu_1=\mu_2=\mu = \hat{\mathbf{k}}\cdot\hat{\mathbf{d}}$. 
The covariance of the Eq.~\ref{eq:Cabi} is now written as

\begin{equation}
    \label{eq:Cabil_pp}
   \begin{split}
        \Cabi &= \int_{0}^{\infty} \frac{k^2 dk}{(2\pi)^3}  \Pabi(k) \\
        &\times\int_\Omega d\Omega \Bi(k, \mu)e^{i\mathbf{k}\cdot\mathbf{r}}\, .
    \end{split}
\end{equation}

Using the Legendre polynomial expansion (Eq.~\ref{eq:LegendreExp}), we obtain

\begin{equation}
    \Bi(k, \mu) = \sum_{\ell^\prime}  \frac{(2\ell^\prime + 1)}{2} L_{\ell^\prime}(\mu) \int_{-1}^{1}d\mu^\prime \Bi(k, \mu^\prime)L_{\ell^\prime}(\mu^\prime)\, .
\end{equation}

Similarly to the wide-angle case, we define the coefficients $\Ml(k)$ as

\begin{equation}
    \Ml(k) = \frac{1}{2}\int_{-1}^{1}d\mu^\prime \Bi(k, \mu^\prime)L_{\ell^\prime}(\mu^\prime)\, .
\end{equation}

In addition, we use the plane-wave expansion (Eq.~\ref{eq:planewave}) such that the Eq.~\ref{eq:Cabil_pp} becomes 

\begin{equation}
\label{eq:Cabil2}
     \begin{split}
       \Cabi &= \int_{0}^{\infty} \frac{k^2 dk}{(2\pi)^3} \Pabi(k) \\ 
       &\times \sum_{\ell, \ell^\prime} i^\ell j_\ell(kr)(2\ell + 1)(2\ell^\prime+1) \Ml(k)\\
      &\times \int_\Omega d\Omega \Lg{}(\normv{k}\cdot\normv{r})L_{\ell^\prime}(\mu).
    \end{split}
\end{equation}

Using the result of Eq.~\ref{eq:int2L} for the angular integral in Eq.~\ref{eq:Cabil2}, considering as a reminder that $\mu = \normv{k}\cdot\normv{d}$, we get

\begin{equation}
    \begin{split}
       \Cabi &= \sum_{\ell} 4\pi\sum_{m=-\ell}^\ell\Ylm{}{}(\normv{d})\Ylms{}{}(\normv{r}) \\
       &\times\int_{0}^{\infty} i^\ell \frac{k^2 dk}{2\pi^2} \Pabi(k)\Mlnp(k)j_\ell(kr)\, .
    \end{split}
\end{equation}

We define the geometrical term $N_{\mathrm{ab},i}^{\ell}(\normv{r},\normv{d})$, and simplify it using the spherical harmonic addition theorem (Eq. \ref{eq:spherical_harmonic}):

\begin{equation}
    N_{\mathrm{ab},\ell}(\normv{r},\normv{d}) = 4\pi\sum_{m=-\ell}^\ell\Ylm{}{}(\normv{d})\Ylms{}{}(\normv{r})=(2\ell +1)L_\ell(\cos\left(\pi - \phi\right))\, .
\end{equation}

In this plane-parallel case, those geometric terms are dependent only on the $\phi$ angle, and thus also on the definition of the referential line-of-sight $\mathbf{d}$. The final expression for the covariance term in the plane-parallel model is given by

\begin{equation}
\label{eq:cov_flip_pp}
\Cabi = \sum_{\ell} N_{\mathrm{ab},\ell}(\phi)\mathcal{H}_\ell\left[\Pabi(k)\Mlnp(k)\right](r)\, .
\end{equation}

This equation corresponds to a similar form to the wide-angle case (equation~\ref{eq:cov_flip_wa}) with simpler analytical integrations. The properties concerning the algorithmic efficiency are still valid. Our framework gives the advantage to include wide-angle effects, which can be very computationally intensive and mathematically complex, in the same formalism.

\section{Likelihood-based field-level inference}
\label{sec:mle}

\begin{figure*}
    \centering
	\includegraphics[width=0.85\textwidth]{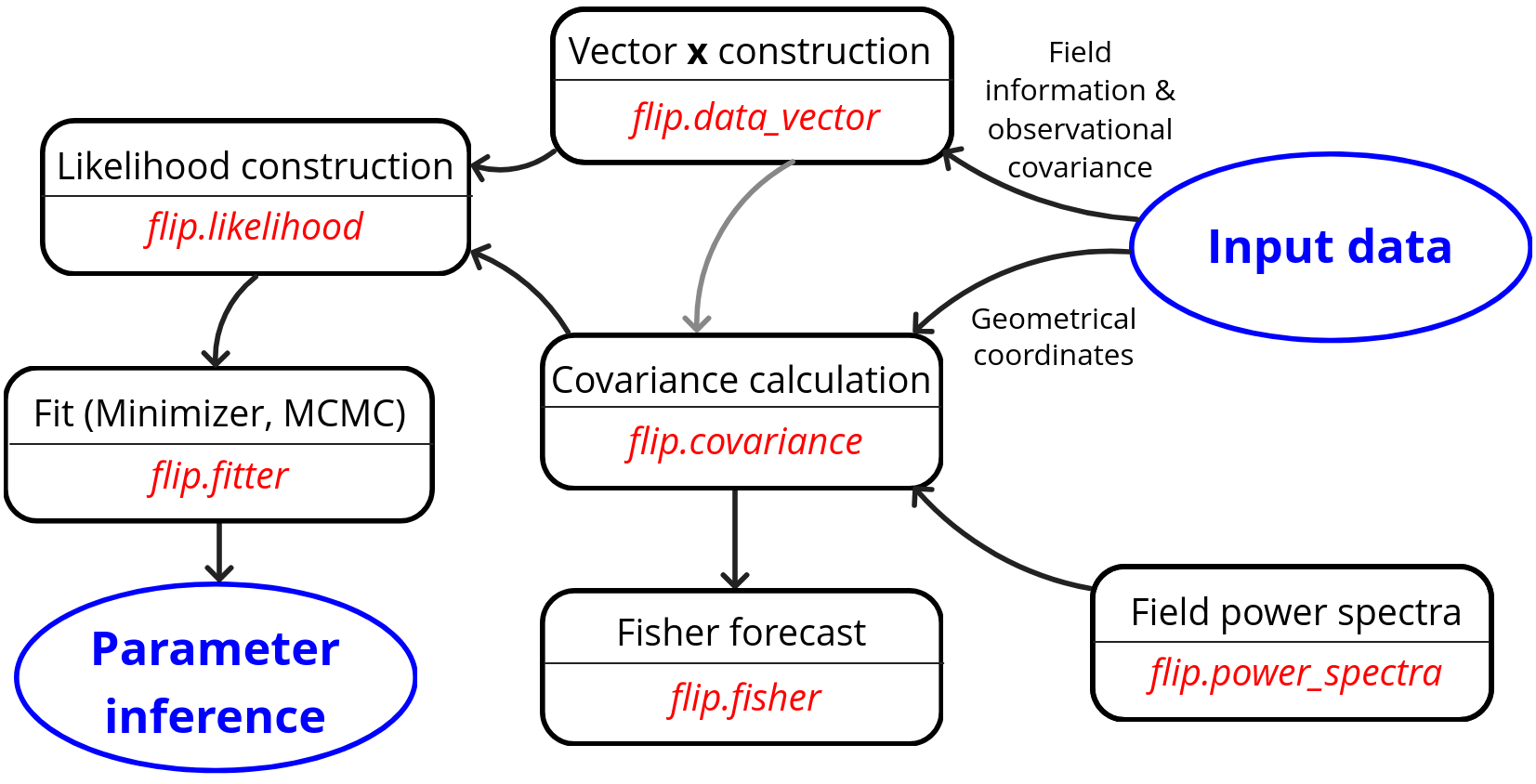}
    \caption{Schematical implementation of the different modules in the \texttt{flip} package. The arrow represents data flows between two modules. The link in gray between vector construction and the covariance calculation represents an alternative way to compute the covariance matrix directly from the data vector object.}
    \label{fig:diagram}
\end{figure*}

The first direct application of the previously derived framework is the inference of cosmological parameters with a likelihood maximization. This method allows one to make an inference on a set of parameters that we note $\Theta$ by maximizing the likelihood whose general form is $\mathcal{L}\left[x(\Theta),C_{xx}(\Theta) , C_{\mathrm{obs}}(\Theta)\right]$, where the field $x$, the field covariance $C_{xx}$, and the observational covariance $C_{\mathrm{obs}}$ depend on a set of parameters $\Theta$. For example, the latter matrix can contain the field error bar or be a full covariance comprising additional field correlations caused by spurious instrumental effects.

The schematic implementation of this method in the \texttt{flip} package is represented in Fig.~\ref{fig:diagram}. We use the input data to generate the field covariance components $\Cabi$, following the derivation from the section~\ref{sec:covariance}, and to create a data vector $x$ formatted in a class which also contains the observational covariance. The likelihood is built from the theoretical covariance and the data vector. The parameters are fitted from this likelihood either by best-fit minimizer or with an MCMC sampling.

\subsection{Covariance model developed for inference}
\label{subsec:covariance}

To better assess the performance of our generalized field covariance framework, we first aimed at reproducing the latest models available in the literature~\citep{adams_improving_2017,adams_joint_2020,lai_using_2022,carreres_growth-rate_2023}. In the second step, we used lessons learned to develop a new covariance model for densities and velocities, accounting for wide-angle effects and with superior performance.

As an illustrative example, we start by decomposing a model used in literature for velocity correlations with wide-angle modeling in~\cite{adams_improving_2017,lai_using_2022,carreres_growth-rate_2023}. Their velocity power spectrum model contains one term:

\begin{equation}
    P_{\mathrm{vv}} = (aHf\sigma_8)^2 \frac{\mu_1 \mu_2}{k^2} P_{\mathrm{\theta\theta}}(k) D_u^2(k, \sigma_u)\, ,
\end{equation}

\noindent where the $a$ is the scale factor, $H$ the Hubble parameter, $f$ the growth rate of structure, $\sigma_8$ the amplitude of the matter perturbations in spheres of $8\ \mpc$ comoving radius, $P_{\mathrm{\theta\theta}}$ is the velocity dispersion power spectrum normalized by fiducial $\sigma_8^2$ value, and $D_u$ is an empirical damping function often used in peculiar velocity studies to model the effect of RSD on the position of velocities itself (see e.g., \cite{koda_are_2014}). In the \texttt{flip} framework of~\cite{carreres_growth-rate_2023} model, we decompose the equation~\ref{eq:Pwflip} over one covariance term:

\begin{align}
    & \Avv = (f\sigma_8)^2\, , \\
    & \Bvv(k, \mu_1, \mu_2) = (aH)^2 \frac{\mu_1 \mu_2}{k^2}\, , \\
    & \mathcal{P}_{\mathrm{vv},0} =  P_{\mathrm{\theta\theta}}(k) D_u^2(k, \sigma_u)\, .
\end{align}

A simple \texttt{sympy} \git{sympy/sympy}{}\citep{sympy} computation of the $\Nlll$ and $\Mll$ terms (in equations \ref{eq:nlldef} and \ref{eq:mlldef}) states that only following terms are nonzero:

\begin{align}
    &M_{\mathrm{vv},n=0}^{\ell_1=1, \ell_2=1} = \frac{(aH)^2}{9k^2}\, , \\
    &N_{\mathrm{vv},l=0}^{\ell_1=1, \ell_2=1} = 3\cos(\alpha)\, , \\
    &N_{\mathrm{vv},l=2}^{\ell_1=1, \ell_2=1} = \frac{( 9 \cos{2 \phi} + 3\cos{\alpha})}{2k^2}\, .
\end{align}

Adding up the different terms to compute the field covariance matrix, we obtain

\begin{align}
\begin{split}
    C_{\mathrm{vv}}(\mathbf{r_1}, \mathbf{r_2}) =& \frac{(aHf\sigma_8)^2}{2 \pi^2} \int_k dk  P_{\mathrm{\theta\theta}} D_u^2(k, \sigma_u) \\
    &\left[ j_0(kr) \frac{\cos(\alpha)}{3}  -  j_2(kr) \frac{( 3 \cos{2 \phi} + \cos{\alpha})}{6}  \right]\, .
\end{split}
\end{align}

Following the demonstration of~\cite{lai_using_2022} in Appendix B3, we obtain the same formula in previous articles~\citep{adams_improving_2017,lai_using_2022,carreres_growth-rate_2023}. Note that the demonstration uses the bisector theorem, meaning that the mathematical equivalency is only valid when the line-of-sight reference is the bisector. The \texttt{flip} covariance framework is unnecessarily complex for simple linear velocity modeling. However, this is not the case for multiple-term power spectrum models with complex forms, which is the case for density models, especially considering nonlinear corrections.

Table \ref{tab:models} lists the terms used in the \texttt{flip} framework to reproduce all the previously designed models and a new model we developed for this study. \cite{adams_improving_2017} developed a model (\textbf{AB17} hereafter) aiming at linking density and velocity information with a density model without RSD. Thanks to this simplification, they were able to account for wide-angles. The first inclusion of RSD in the density terms was realized in the model presented in~\cite{adams_joint_2020}, named \textbf{AB20}. As a simplification, they choose to perform the integration for a plane parallel model, i.e., taking the RSD anisotropic power spectra for $vv$, $gv$, and $vv$ terms. This assumption breaks for large-area surveys at low redshift, and it is thus insufficient to properly model the clustering for the current and future generations of survey. However, this plane-parallel model is adapted to test the wide-angle models in the limit of very high redshift. For the case of~\cite{carreres_growth-rate_2023}, named \textbf{C23}, which only contains a velocity modeling $vv$, the terms correspond to the wide-angle $vv$ of \textbf{AB17}. The authors in \cite{lai_using_2022} propose a wide-angle model, which we call \textbf{L22}, by performing a Taylor expansion on the RSD finger-of-god (FoG) small-scale term:

\begin{equation}
    e^{-\frac{\left(k \mu \sigma_{\mathrm{g}}\right)^2}{2}}=\sum_{p=0}^{\infty} \frac{(-1)^p\left(k \sigma_{\mathrm{g}}\right)^{2 p}}{2^p p!} \mu^{2 p}\, .
\end{equation} 

Since this decomposition is performed two times for the $gg$ terms, one time for the $gv$ terms, and calculated to the $p=4$ order, it yields many covariance terms. In table~\ref{tab:models}, we have decomposed this model in the \texttt{flip} formalism by summing the density-density ($gg$) terms which have the same value for the sum $m = p + q$. For the $gv$ cross-correlation terms, we have the same formalism as \textbf{L22}, renaming the $p$ index by $m$ for consistency. This model allows ont to directly fit the FoG parameter ($\sigma_{\mathrm{g}}$) and manages to express all the covariance computations in terms of Hankel transforms, thus speeding up the calculation. The main weaknesses of this model are that the Taylor expansion is computed up to a certain fixed order and that it is based on the assumption that the product $k \mu \sigma_{\mathrm{g}}$ is small compared to unity. Knowing that $\sigma_{\mathrm{g}}$ values are generally considered ranging $[1 - 10]\ \mpc$, it means that the model is only valid for low values of $k$ (typically $k < 0.1\ h.\mathrm{Mpc}^{-1}$). It is limiting, especially for an extension to nonlinear models which needs small-scale modes for integration stability, directly at the power spectrum level ($\Pabi$) or in the full power spectrum model ($P_{\mathrm{ab}}(k, \mu_1, \mu_2)$). The authors of the \textbf{L22} model decided to separate small-scale modeling ($ 0.2 < k < 1.0\ h.\mathrm{Mpc}^{-1}$) with the rest of the scales considered and introduce a dedicated nuisance parameter for the amplitude of the covariance in the small scales.

In this study, we created the new covariance model, denoted as \textbf{RC25}, which frees itself from the Taylor expansion approximation while keeping the wide-angle modeling. Removing the Taylor expansion allows one to extend modeling to larger wavenumbers ($k > 0.1\ h.\mathrm{Mpc}^{-1}$), which is necessary for the use of nonlinear power spectrum models. As shown in table~\ref{tab:models}, we obtain this model by reverting the \textbf{L22} model Taylor-expansion, or equivalently by including wide-angle terms in the \textbf{AB20} model. Contrary to the \textbf{L22} model, \textbf{RC25} cannot account for $\sigma_{\mathrm{g}}$ nuisance parameter directly in the linear decomposition of the power spectrum model. We overcome this issue 
by pre-computing covariances for several values of $\sigma_{\mathrm{g}}$ and interpolating during the inference. This interpolation procedure will be detailed in section~\ref{subsec:like_vec}.

The \texttt{flip.covariance} module of the figure~\ref{fig:diagram} contains for all the models previously described, the terms $\Nlll(\phi,\alpha)$ and $\Mll(k)$ terms (respectively, $N_{\mathrm{ab},\ell}(\phi)$ and $\Mlnp(k)$ for plane-parallel models). The integration inside those terms is performed analytically with the symbolic library \texttt{sympy}. The derived functions are stored inside the \texttt{flip.covariance} module by symbolic code generation.

When a covariance model is generated following the equation~\ref{eq:cov_flip_wa} (or~\ref{eq:cov_flip_pp} in the plane-parallel case), the geometrical terms $\Nlll(\phi,\alpha)$ ($N_{\mathrm{ab},\ell}(\phi)$) are computed for all the field pairs considered. In a second step, the term $\mathcal{H}_\ell\left[\Pabi(k) \Mll(k)\right]$ is computed numerically by first expressing $\Mll(k)$ (or $\Mlnp(k)$ in the plane-parallel case) for the considered pair and by computing the Hankel transform $\mathcal{H}_\ell$. This last step is a numerical integration performed over the wavenumber $k$ and is accelerated by employing the FFTLog algorithm~\citep{talman1978}. For a given data field vector $x$ of size $N$, we must compute $N(N+1)/2$ times this equation. In order to speed up this process, we distribute the computation of the covariance with multi-threading by separating the covariance into arrays.

When the correlation is computed between two different fields, for example the radial velocity, $v$, and galaxy density, $\delta_{\mathrm{g}}$, scalar field, we only compute $C_{gv}$ and deduce $C_{vg}$. \cite{adams_improving_2017} (Appendix A3) showed that a minus sign appears between $C_{vg}$ and $C_{gv}$ when employing their definitions. However, the equation for cross terms is not symmetric because only odd Legendre polynomials are nonzero, so changing the $\mathbf{r}$ vector between $gv$ and $vg$ correlations introduces another minus sign. Consequently, we can simply take $C_{vg}$ as the transpose of $C_{gv}$.

Some numerical considerations must be considered to compute all the covariance models shown in table~\ref{tab:models}. For models including RSD FoG terms without Taylor expansion such as \textbf{RC25} and \textbf{AB20}, the $\Mll(k)$ or $\Mlnp(k)$ terms can lead to numerical instabilities due to linear combination of large floats. A numerical correction, detailed in the appendix~\ref{appendix:numerical}, allows us to solve this issue. Additionally, the use of the Hankel transform can introduce numerical instabilities. The latter can be caused by considering the smaller separations between the pair for which we want to compute the covariance. In the special case of simulations, other instabilities can occur at large scales when the numerical integration over wavenumber is performed on a range not covered by the considered simulation. We solve this issue by introducing a regularization term for low wavenumbers, also detailed in the appendix~\ref{appendix:numerical}.

\subsection{Vector and Likelihood estimators}
\label{subsec:like_vec}

\begingroup
\renewcommand{\arraystretch}{2}
\begin{table*}[]
    \centering
    \caption{Velocity redshift dependence implemented in \texttt{flip} to convert Hubble diagram residuals into peculiar velocities.}
    \resizebox{0.9\textwidth}{!}{\begin{tabular}{c||c|c|c|c}
         \hline\hline
         Name            & Watkins               &                  Low-$z$ &           Hubble high-order                     & Full\\
         \hline
        $p_i$ & $z_{\mathrm{obs},i}$ & $z_{\mathrm{obs},i}$ & $z_{\mathrm{obs},i}$, $q_0$, $j_0$ & $z_{\mathrm{obs},i}$,$D(z_{\mathrm{obs},i})$, $H(z_{\mathrm{obs},i}) / h$ \\
        \hline
        $J(p_i)$ & $\frac{z_{\mathrm{obs},i}}{1+z_{\mathrm{obs},i}}$ & $ z_{\mathrm{obs},i}$ & $\frac{z_{\mathrm{obs},i} \left( 1  + (1/2) (1 - q_0) z_{\mathrm{obs},i} - (1/6) (1 - q_0 - 3 q_0^2 + j_0) z_{\mathrm{obs},i}^2\right)}{1+z_{\mathrm{obs},i}}$ & $ \left(\frac{(1+z_{\mathrm{obs},i})c}{(H(z_{\mathrm{obs},i})/h) D(z_{\mathrm{obs},i})} - 1 \right)^{-1}$ \\
        \hline\hline
    \end{tabular}}
    \tablefoot{In addition to the terms defined in the text, the $q_0$ and $j_0$ parameters are, respectively, the deceleration and jerk parameters in the higher-order development of the Hubble law.}
    \label{tab:velest}
\end{table*}
\endgroup

As shown in Fig.~\ref{fig:diagram}, the \texttt{flip} package includes modules to perform cosmological parameter fits. The input data are handled by different Python classes implemented in the \texttt{flip.data\_vector} module. This module includes various Python classes to handle density and velocity survey data. 

For peculiar velocities, several estimators are implemented to cover different types of surveys, such as TF, FP, and \sn. For simulation, where true peculiar velocity is accessible, a data vector mode with the direct velocities is implemented, with an option for grouping them when they are near each other. For TF and FP studies, the \texttt{flip} software contains a commonly used velocity estimator for a galaxy $i$ based on the logarithmic distance ratio:

\begin{equation}
    \label{eq:etatovel}
    v_i =\ln (10) \frac{D\left(z_{\mathrm{obs},i}\right) H\left(z_{\mathrm{obs},i}\right)}{\left(1+z_{\mathrm{obs},i}\right)} \eta_i\, ,
\end{equation}

\noindent where $\eta_{i} =\log \left[D\left(z_{\mathrm{obs},i}\right)/D\left(z_{\mathrm{cos},i}\right)\right]$ is the logarithmic distance ratio of the galaxy $i$, $z_{\mathrm{obs},i}$ and $z_{\mathrm{cos},i}$ are, respectively, the observed and cosmological redshifts, and $D$ is the comoving radial distance expressed in $\mpc$.

The \texttt{flip} software also contains a method to go from Hubble diagram residuals $\Delta \mu_i$ to peculiar velocities, which can be used for any distance indicator (TF, FP, \sn). Several methods for transforming Hubble diagram residuals to peculiar velocities are implemented; they take the form

\begin{equation}
    \label{eq:dmutovel}
    v_i = -\frac{c \ln(10)}{5} J(p_i) \Delta\mu_i \, ,
\end{equation}

\noindent where $\mathbf{p}$ are the parameters needed to estimate the velocities. The currently implemented estimators are given in Table~\ref{tab:velest}. 

The uncertainty of the radial peculiar velocity estimate $\sigma_{v,i}$ is computed by propagating the estimated error in the observable used. If velocities are directly used, \texttt{flip} needs an estimate of the error on those velocities as an input. For the logarithmic distance (respectively, Hubble residuals), the error bar on velocities is computed by replacing $\eta_i$ (resp.\ $\Delta\mu_i$) on equation~\ref{eq:etatovel} (resp.~\ref{eq:dmutovel}) by the error on logarithmic distance $\sigma_{\eta,i}$ (resp.\ Hubble residual $\sigma_{\Delta\mu,i}$).

For the special case of \sn, we express explicitly the Hubble residuals as a function of light-curve fitting parameters. Those residuals are expressed using the Tripp relation~\citep{Tripp1998} and given in greater detail in~\cite{carreres_growth-rate_2023}:

\begin{equation}
\begin{split}
    \Delta \mu_i (\Theta_{\mathrm{HD}}) &= \mu_{\mathrm{obs},i}(\Theta_{\mathrm{HD}}) - \mu_{\mathrm{model},i} \\
    &= m_{B,i} - M_0 + \alpha x_{1,i} - \beta c_i  \\
    &\quad - \left( 5 \log\left[(1+ z_{\mathrm{obs},i})D(z_{\mathrm{obs},i})\right] + 25  \right)\,,
\end{split}
\end{equation}

\noindent where $m_{B,i}$ is the apparent magnitude of SN~Ia $i$ in the B-band, $M_0$ is the absolute magnitude of \sn~which can be seen as an offset of the SN~Ia Hubble diagram, $\alpha$ and $\beta$ correct for correlations of Hubble residuals with SN~Ia stretch $x_{1,i}$ and color $c_i$ parameters. For this SN~Ia implementation, the uncertainty on Hubble residuals is computed as: 

\begin{equation}
\begin{split}
    \sigma_{\Delta \mu,i}^2(\Theta_{\mathrm{HD}}) = \irow{1&\alpha&-\beta} \cdot C_{\mathrm{lcfit},i} \cdot \irow{1&\alpha&-\beta}^T +\sigma_M^2\, ,
\end{split}
\end{equation}

\noindent where the $C_{\mathrm{lcfit},i}$ is the covariance of the light-curve fit for SN~Ia $i$, and $\sigma_M$ is the SN~Ia intrinsic scatter (assumed to be color-independent). This explicit parameterization allows us to fit simultaneously for SN~Ia Hubble diagram's nuisance parameters $\Theta_{\mathrm{HD}} = \left\{M_0,\alpha,\beta,\sigma_M \right\}$ along with the cosmological parameters of interest. This type of parameterization can be adapted for TF and FP empirical relations, and we plan to develop it in future studies. 

The velocity estimator presented here or in equation~\ref{eq:etatovel} can have an average not necessarily null. It can rise from an actual bulk flow in the cosmic web. In this model, the SN Ia intrinsic magnitude dispersion is also added as a free parameter to the covariance matrix at the fitting stage. We added an option to all velocity estimators to fit for a velocity zero point offset.

The galaxy density field, $\delta_{\mathrm{g}}$, is computed from a galaxy catalog using mesh assignment schemes available in \texttt{pypower} \git{cosmodesi/pypower}{} software. The galaxy density field in a specific cell $i$ is defined by $\delta_{\mathrm{g},i} = \left(N_{\mathrm{cell},i} / N_{\mathrm{exp},i}\right)  - 1$ where  $N_{\mathrm{cell},i}$ is the number of galaxies associated with the cell, $i$, and $N_{\mathrm{exp},i}$ is the expected number. The latter is computed using a random catalog normalized to the same number of galaxies as the main catalog. The number of assigned galaxies, $N_{\mathrm{cell},i}$, can vary depending on the adopted resampling scheme: nearest grid point (NGP), cloud-in-cell (CIC),  triangular-shaped cloud (TSC), or piecewise cubic spline (PCS). This resampling method changes which neighboring cells are associated with a galaxy, effectively changing the smoothing of the density field. The uncertainty associated with the density field is estimated as $\sigma_{\delta,\mathrm{g},i} = 1 / \sqrt{N_{\mathrm{exp},i}}$, considering that $N_{\mathrm{exp},i} \gg 1$. As the assignment scheme suppresses power at small scales, the input power spectra are corrected by multiplying them to the following grid window function:

\begin{equation}
    \label{eq:grid_corr}
    \Gamma(k, L)=\left\langle\frac{8}{L^3} \left(\frac{\sin \left(k_x \frac{L}{2}\right)}{k_x} \frac{\sin \left(k_y \frac{L}{2}\right)}{k_y} \frac{\sin \left(k_z \frac{L}{2}\right)}{k_z} \right)^{n}  \right\rangle_{\mathbf{k}}\, ,
\end{equation}

\noindent where $n$ depends on the resampling scheme chosen (1 for NGP, 2 for CIC, 3 for TSC, and 4 for PCS). 

The likelihood implemented in \texttt{flip} to perform the likelihood-based field-level inference is a multivariate Gaussian function:

\begin{equation}
    \mathcal{L}[x
    (\Theta), C(\Theta)] = (2\pi)^{-N_x/2}|C|^{-\frac{1}{2}} \exp\left[-\frac{1}{2}x^TC^{-1}x\right]\, ,
\end{equation}

\noindent where the covariance matrix used is $C = C_{xx} + C_{\mathrm{obs}}$ where $C_{xx}$ is the theoretical field covariance of the scalar field $x$, and $C_{\mathrm{obs}}$ is an observational covariance matrix. The latter can be a diagonal matrix containing the error bar associated with the field and scaled by a field-dependent parameter $\sigma_{x}$ or the observational covariance matrix provided by a given survey.

Furthermore, the \texttt{flip} package contains an option to interpolate the field covariance matrix when parameters to fit do not factor into a coefficient of the linear decomposition of the model power spectrum in equation~\ref{eq:Pwflip}. For peculiar velocities, we interpolate the covariance matrix depending on the velocity position FoG parameter $\sigma_{\mathrm{u}}$ and include that in the fit. For density models that do not include the FoG parameter $\sigma_{\mathrm{g}}$ in the model power spectrum linear decomposition, such as \textbf{AB20} or \textbf{RC25}, we also interpolate with respect to this parameter. The interpolation can also be performed for two parameters, for example, when the fit is performed jointly for galaxy density $\delta_{\mathrm{g}}$ and velocity $v$ field. For that case, the data vector is given by  
\begin{equation}
    x = \begin{pmatrix}
        \delta_{\mathrm{g}}\\
        v
    \end{pmatrix},
\end{equation}
and the field covariance is organized as:
\begin{equation}
    C_{xx} = \begin{bmatrix}
        C_{\delta\delta} & C_{\delta v} \\
        C_{v \delta}     & C_{vv} 
    \end{bmatrix}\, .
\end{equation}

The shape of the likelihood implies that the field $x$ is considered as Gaussian. This approximation holds for noisy fields and large cosmological scales but generally breaks down on small scales. We plan to extend the likelihood of the \texttt{flip} software to account for the observational effects changing the field's Gaussianity, such as selection effects. Finally, the likelihoods are minimized in \texttt{flip} either with the best-fit minimizer \texttt{iminuit} \git{scikit-hep/iminuit}{}~\citep{minuit1975,iminuit2020}, or by performing a Markov-chain Monte Carlo (MCMC) sampling with the \texttt{emcee} \git{dfm/emcee}{}~\citep{emcee} software. When performing an MCMC sampling, the \texttt{flip} software contains implementations for positive, uniform, and Gaussian priors $P(\Theta)$. The posterior distribution $P(\Theta \mid x)$ which is sampled is then given by the Bayes theorem:

\begin{equation}
P(\Theta \mid x) = \frac{\mathcal{L}[x(\Theta), C(\Theta)] P(\Theta)}{P(x)}\, ,
\end{equation}

\noindent where the evidence $P(x)$ is not varied and used for normalizing the posterior distribution to unity.

The \texttt{flip} software also contains a module for isotropic power spectrum generation for velocity divergence and matter. Those power spectrum can be generated using the \texttt{CLASS} \git{lesgourg/class_public}{}~\citep{class2011} Boltzmann solver, and velocity dispersion models are generated following the \cite{bel_accurate_2019} models.

\section{Validation of the \texttt{flip} software}
\label{sec:validation}

The main objective of this section is to show the additional applications of the \texttt{flip} software while validating the field covariance models we implemented. In particular, we aim at comparing \texttt{flip} to previous codes, comparing the covariance models between each other for a fixed regular grid of coordinates, and validating those models with an estimation of the two-point correlation function on an N-body simulation.

\subsection{Comparison with previous code}

In previous SN~Ia studies, the velocity covariance matrix $C_{vv}$ was generally calculated using the code \texttt{pairV} \citep{hui_corr_2006}. This code was created to show that the coherent motion caused by large-scale structure is in fact important for parameter estimation in SN~Ia cosmology, and as such calculates the coherent-motion-induced magnitude covariance between arbitrary points in space, as demonstrated by \citet{davis_effect_2011}. The \texttt{pairV} code can therefore be used as the standard to compare with \texttt{flip}, numerically and performance-wise. To demonstrate this, we generate a mock set of positions mimicking a SN~Ia survey at a range of distances and angular positions and calculate the covariance using each code. The sky positions were chosen to have uniform random angular separations from 0$^{\circ}$ to 180$^{\circ}$, and the redshifts were normally distributed with a mean $z=0.025$ and a standard deviation of $z=0.025$ (with a minimum distance cut of 10 Mpc) to approximately reproduce a low-$z$ survey.

\begin{figure}
    \centering
	\includegraphics[width=0.95\columnwidth]{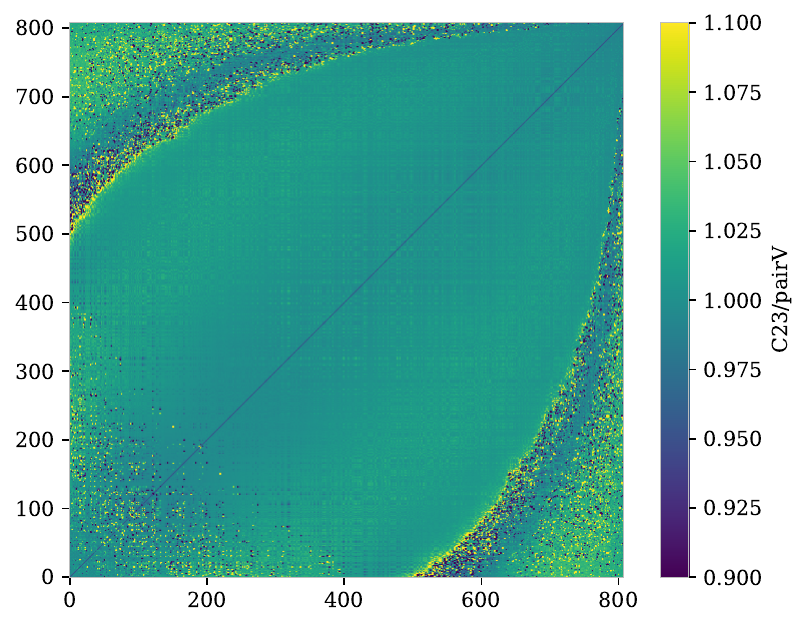}
    \caption{Ratio of $C_{vv}$ as calculated using the \textbf{C23} model (equivalent to \textbf{RC25} velocity) and the \texttt{pairV} for a mock set of 3D positions using the same cosmology. The agreement is on average very close, but there are regions with instabilities.}
    \label{fig:flip_pairV_cov_comp}
\end{figure}

The ratio of these matrices is shown in figure~\ref{fig:flip_pairV_cov_comp}. We used the same cosmology \citep[that of][]{carreres_growth-rate_2023} for both and the \textbf{C23} model (equivalent to \textbf{RC25} velocity) in \texttt{flip}, and converted them to be in the same units (arbitrarily, \texttt{flip} to magnitude-space). We used the native power spectrum for each code, which we chose to be fully linear for this comparison. For display purposes, the data are ordered by redshift to show the dependence on separation. For ease of comparison between the codes, we created a Python version of \texttt{pairV} called \texttt{pypairV} \git{dparkins/pypairV}{} in which we have only updated the original code from FORTRAN77 to Python and validated that they give precisely the same results.

\begin{figure}
    \centering
	\includegraphics[width=0.95\columnwidth]{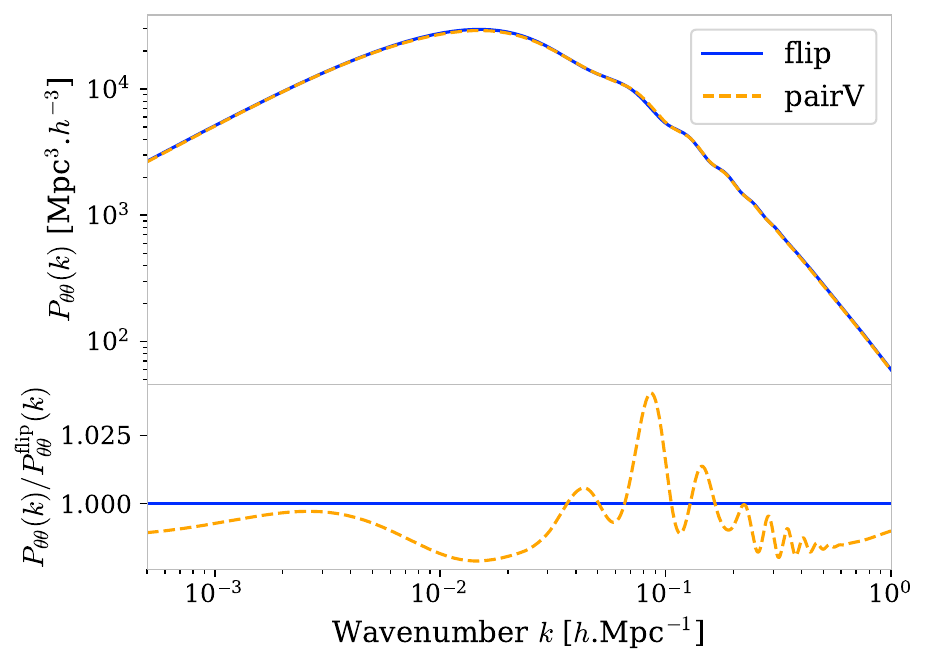}
    \caption{Comparison of the linear power spectra generated using \texttt{flip} and \texttt{pairV}. Whereas \texttt{flip} makes use of the \texttt{CLASS} Boltzmann code, \texttt{pairV} uses a fitting formula from \cite{EH-transfer_1998}. The differences at the $P_{\theta\theta}$ level cause similar differences at the covariance level.} 
    \label{fig:powspec_comp}
\end{figure}

The agreement between the codes is excellent, but the different choices in each code cause some structured residuals. The two main ones are the power spectrum generation and the covariance integration method. The \texttt{pairV} code opted to approximate the power spectrum with the transfer function developed by \citet{EH-transfer_1998} that very closely reproduces the full Boltzmann code calculation, by for example~\texttt{CLASS}, used in \texttt{flip}.  Both are shown in figure~\ref{fig:powspec_comp}. The percent-level differences between the power spectra also cause percent-level gradients in the covariance ratio. 

The most striking feature, however, is the numerical instability in the residual plot (figure~\ref{fig:flip_pairV_cov_comp}). This is caused primarily by the method \texttt{pairV} uses to decompose the integral over the power spectrum. Neglecting prefactors and focusing on the power spectrum integration, the velocity covariance can be written as the addition of a parallel component to the line-of-sight, $\Pi(r)$, and perpendicular, $\Sigma(r)$, as \citep{gorski_1988, gordon_2007, davis_effect_2011},

\begin{equation}
    C^{ij}_{vv} \propto (\hat{\mathbf{r}}_{\mathbf{1}}\cdot \hat{\mathbf{r}})(\hat{\mathbf{r}}_{\mathbf{2}}\cdot \hat{\mathbf{r}})\Pi(r) + \left(\hat{\mathbf{r}}_{\mathbf{1}}\cdot \hat{\mathbf{r}}_{\mathbf{2}}-(\hat{\mathbf{r}}_{\mathbf{1}}\cdot \hat{\mathbf{r}})(\hat{\mathbf{r}}_{\mathbf{2}}\cdot \hat{\mathbf{r}})\right)\Sigma(r)\, ,
\end{equation}

\noindent where the vectors are those depicted in figure~\ref{fig:angles}. Focusing on the integration part, the parallel and perpendicular components are defined as

\begin{equation}
    \Pi(r) \propto \int_k \frac{dk}{2\pi^2}P_{\theta\theta}\left(j_0(kr)-\frac{2j_1(kr)}{kr}\right)\, ,
\end{equation}

\noindent and

\begin{equation}
    \Sigma(r) \propto \int_k \frac{dk}{2\pi^2}P_{\theta\theta}j_1(kr)\, .
\end{equation}
The components are pre-computed and then interpolated within the \texttt{pairV} code, which enables the speed-up compared to the direct, observer-centric approach \citep[for the proof of equivalence of both forms, see][]{davis_effect_2011}. The separation-centric form is used due to its computation speed; while \texttt{pairV} also contains the observer-centric form, it becomes unfeasible to use beyond even a couple of hundred \sn~because of a large increase in the CPU time requirement. However, the same calculation can be performed by \texttt{flip} on the order of seconds, rather than \texttt{pairV}'s many hours.

\begin{figure}
    \centering
	\includegraphics[width=0.95\columnwidth]{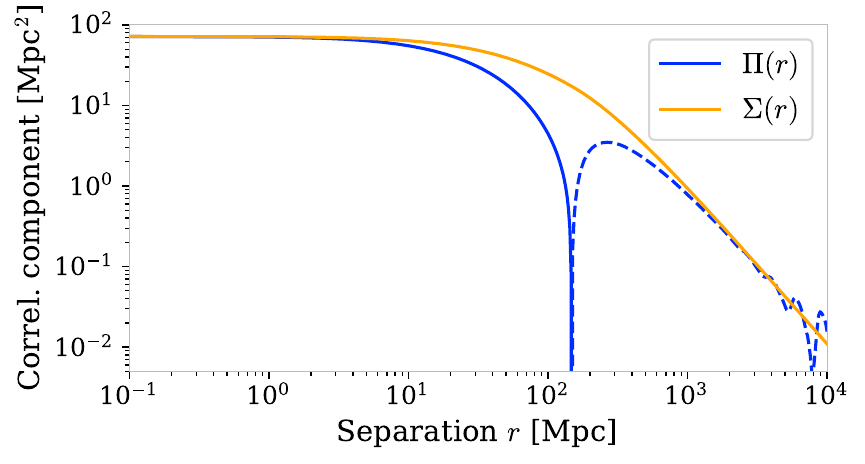}
    \caption{Parallel ($\Pi$, solid blue line) and perpendicular ($\Sigma$, solid orange line) correlation components as a function of physical separation. Since $\Pi(r)$ turns negative at around 150 Mpc, we plot $-\Pi(r)$ with a dashed blue line.} 
    \label{fig:Pi_Sig_funcs}
\end{figure}

Figure~\ref{fig:Pi_Sig_funcs} shows the decomposition components as a function of separation, where it can be seen that $\Pi(r)$ becomes numerically unstable at large separations and also turns negative at around 150 Mpc for the chosen cosmology. As a direct consequence, this is the same scale above which the numerical instabilities appear in figure~\ref{fig:flip_pairV_cov_comp}; the curves of high instability delineate the regions where separations are always larger than 150 Mpc (upper left and lower right) and always below 150 Mpc (lower left). In addition to this source of instability, there is also the fact that the covariance at large separation is so small that tiny differences cause large ratios. At low separations (lower left region) the cause of instability is not as clear, although it improves notably when using identical power spectra between codes. We conclude that \texttt{flip} calculates an almost identical velocity field covariance matrix as the standard \texttt{pairV} code on the range of scales where the integration is performed similarly, and any differences are ultimately negligible for cosmology.

\subsection{Covariance contraction}
\label{subsec:contraction}

\begin{figure*}
    \centering
	\includegraphics[width=0.9\textwidth]{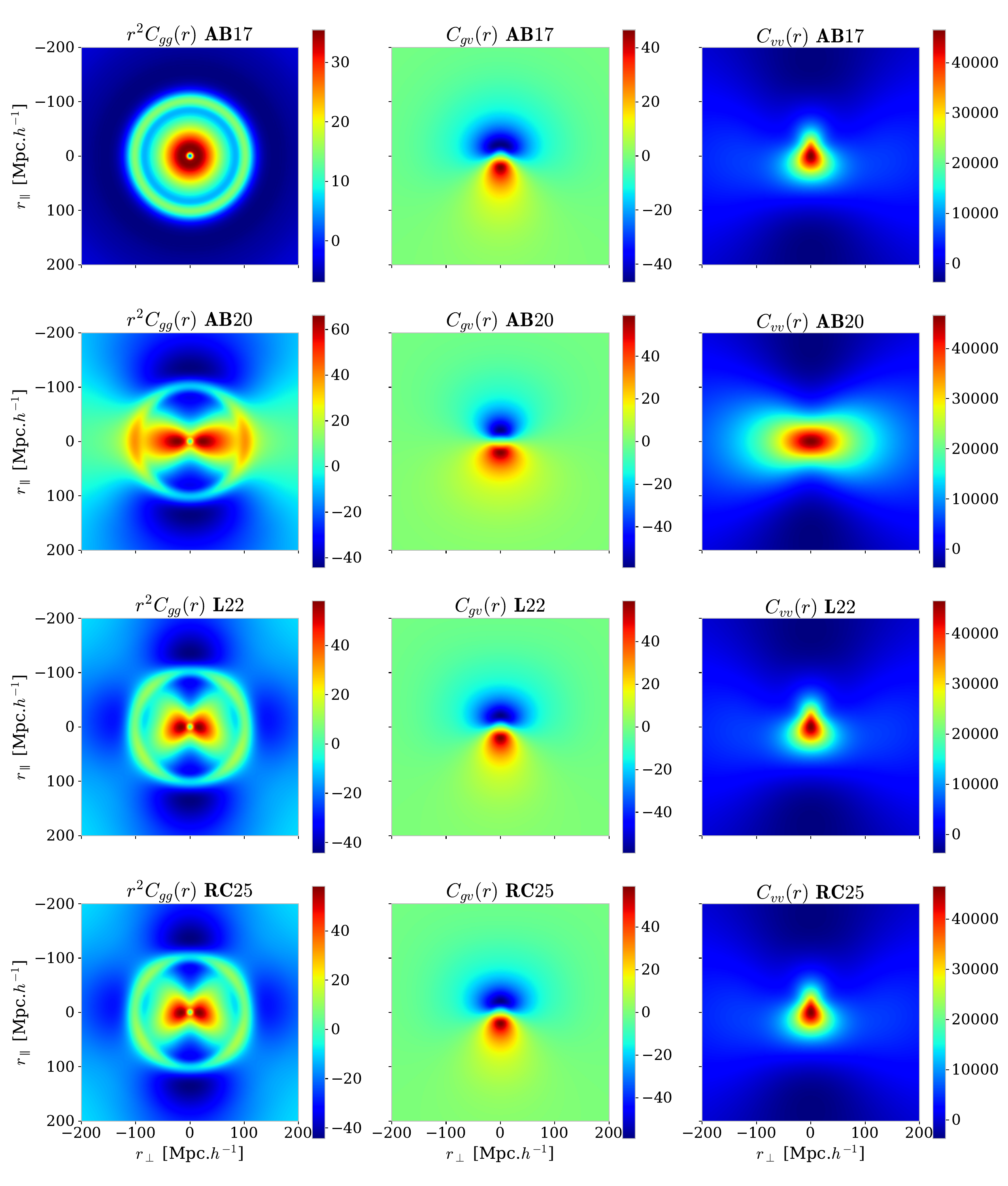}
    \caption{Expression of the field covariance for a uniform coordinate grid, i.e. covariance contraction, for all the models implemented in \texttt{flip}. From top to bottom the \textbf{AB17}, \textbf{AB20}, \textbf{L22}, and \textbf{RC25} models are represented (see table~\ref{tab:models} for reference). The \textbf{C23} model is fully equivalent to $vv$ contractions in the \textbf{AB17}, \textbf{L22}, and \textbf{RC25} models. From left to right are shown the density-density ($gg$), density-velocity ($gv$) and velocity-velocity ($vv$) covariance contractions as a function of transverse ($r_{\bot}$) and line-of-sight ($r_{\parallel}$) separations. The $gg$ contraction is multiplied by the squared separation to highlight the BAO pattern.}
    \label{fig:contraction}
\end{figure*}

\begin{figure*}
	\includegraphics[width=0.95\textwidth]{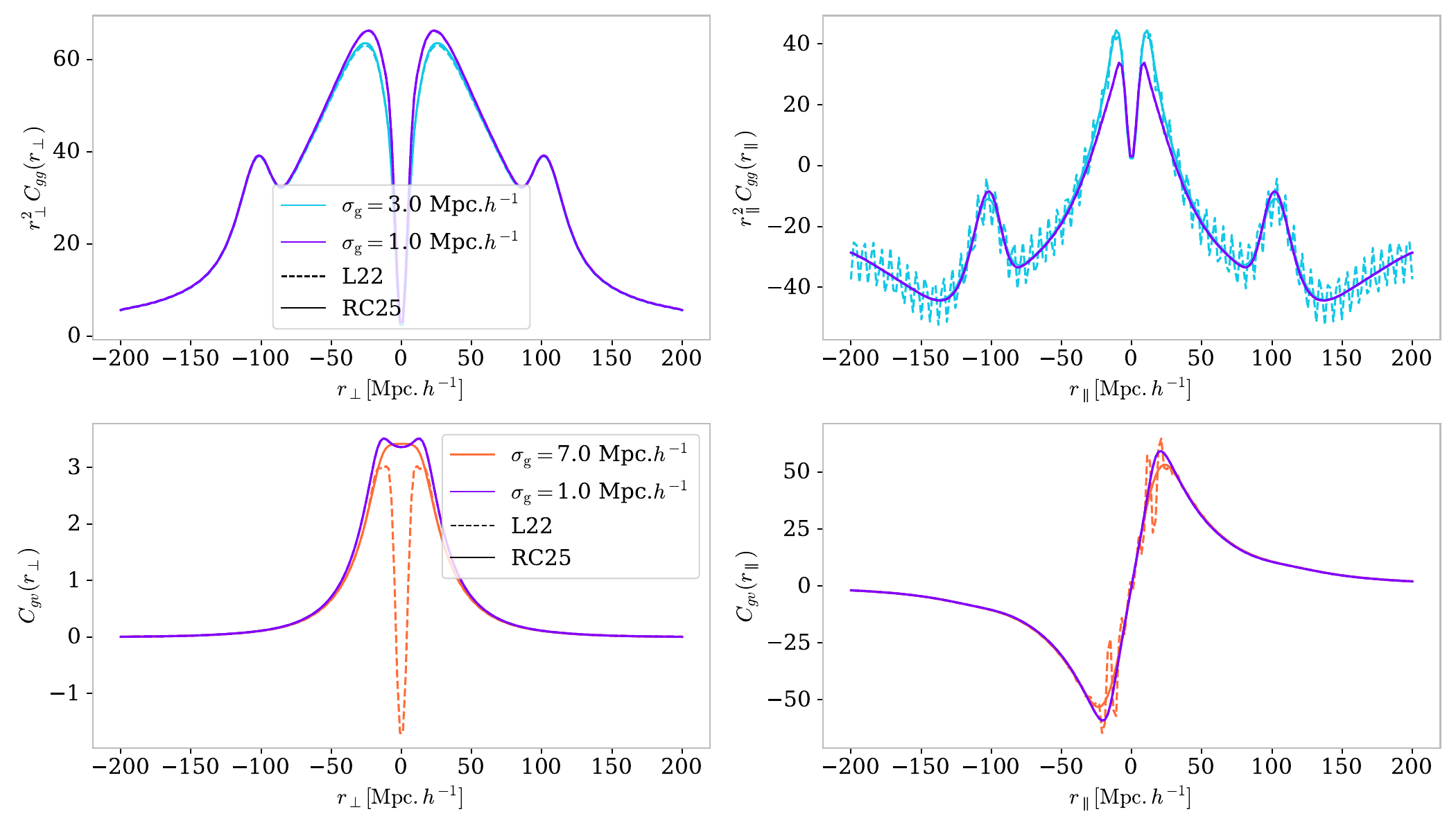}
    \caption{Covariance contraction with the same parameterization than \ref{fig:contraction} while varying the $\sigma_{\mathrm{g}}$ parameter for \textbf{L22} and \textbf{RC25} models. Only $C_{gg}$ (top) and $C_{gv}$ (bottom) are represented, as velocity-velocity correlations do not depend on $\sigma_{\mathrm{g}}$. Those contractions are expressed in one-dimension as a function of $r_{\bot}$ when $r_{\parallel} = 0\ \mpc$ (left) and as a function of $r_{\parallel}$ when $r_{\bot} = 0\ \mpc$ (right).}
    \label{fig:lai_instabilities}
\end{figure*}

\begin{figure}
	\includegraphics[width=0.95\columnwidth]{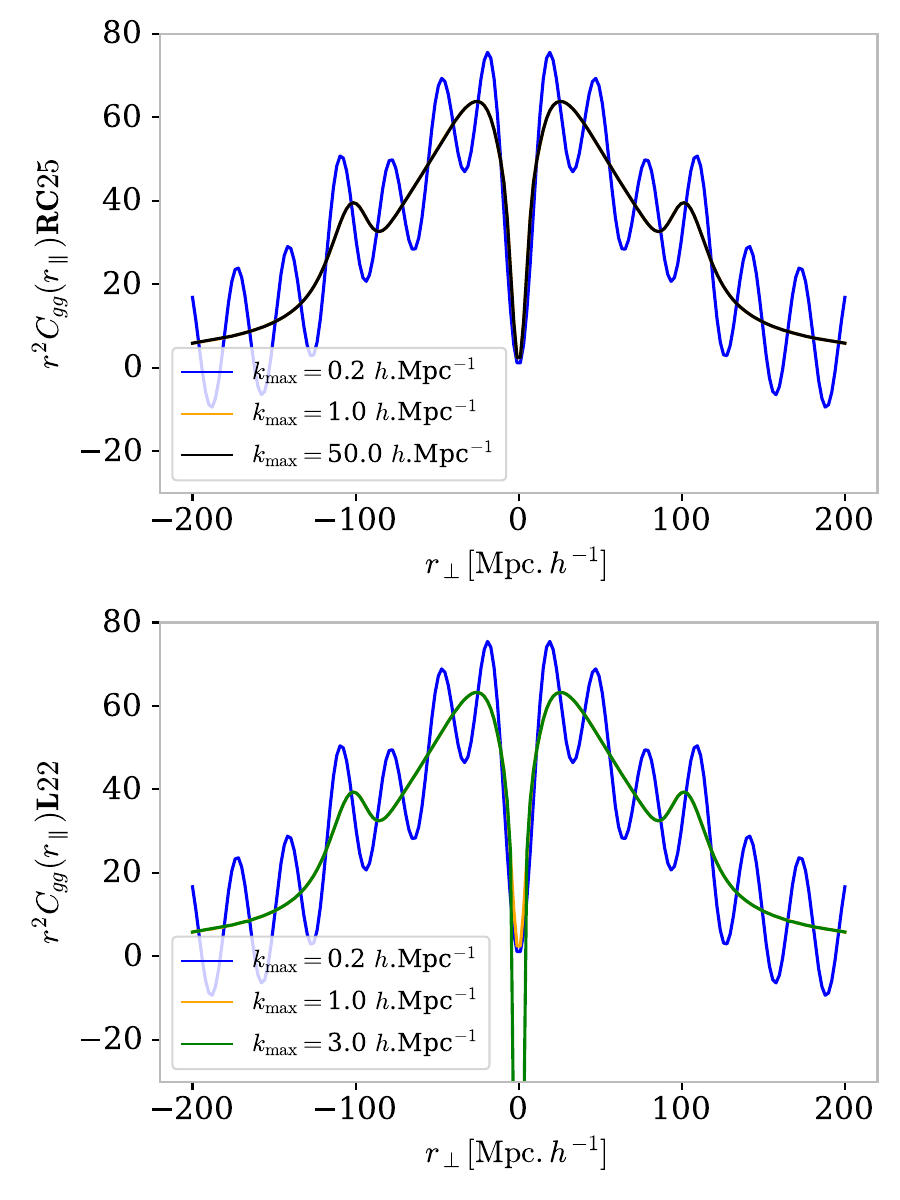}
    \caption{Covariance contraction for \textbf{RC25} (top) and \textbf{L22} (bottom) models with the same parameterization than \ref{fig:lai_instabilities}, and $\sigma_{\mathrm{g}} = 3.0\ \mpc$, while varying the maximal wavenumber used in the integration and denoted here as $k_{\mathrm{max}}$. Those contractions are expressed in one dimension as a function of $r_{\bot}$ when $r_{\parallel} = 0\ \mpc$. For \textbf{RC25} the $k_{\mathrm{max}} = 1.0\ h.\mathrm{Mpc}^{-1}$ and $k_{\mathrm{max}} = 50.0\ h.\mathrm{Mpc}^{-1}$ lines are superimposed.}
    \label{fig:lai_instabilities2}
\end{figure}

All the field covariance models in table~\ref{tab:models} are integrated consistently in the \texttt{flip} framework, making it possible to perform a model comparison. By expressing the field covariance matrix on a fixed regular coordinate grid, a step that we call contraction, we can express a model giving a theoretical estimator of the two-point correlation function.

We link the coordinates defined in figure~\ref{fig:angles} to a basis commonly used in the two-point correlation modeling, i.e., line-of-sight and transverse separation $(r_{\bot}, r_{\parallel})$ or $(r,\mu)$. We define this basis with respect to the vector $\mathbf{d}$ so that the $r = \hat{\mathbf{r}}$ and $\mu = \hat{\mathbf{k}}\cdot\hat{\mathbf{d}}$ definition are equivalent to the one we previously defined. The coordinates $r_{\bot}$ and $r_{\parallel}$ are defined as transverse and parallel separation such that $r_{\parallel} = r \cos(\phi)= r \mu$ and $r_{\bot} = r \sin(\phi)$.

In order to compute the field covariance model, we need to express the $(r, \phi, \alpha)$ parameters in the $(r_{\bot}, r_{\parallel})$ or $(r,\mu)$ basis. Considering the previous basis definitions, only the $\alpha$ parameter cannot be directly expressed. Consequently, for wide-angle models, which explicitly depend on $\alpha$, we need to provide additional information to define this angle. It can be done by fixing $\mathbf{r_1}$ to a specific value which becomes a reference for expressing the strength of wide-angle correction. The impact of the $\mathbf{r_1}$ reference position is shown in more detail in appendix~\ref{appendix:waimpact}. Fixing this reference, we can express the $\alpha$ angle depending on the definition of the line-of-sight reference $\mathbf{d}$ (see equation \ref{eq:d_def}), such that:

\begin{equation}
    \alpha_{\mathrm{bisector}} = 2 \arcsin\left(\frac{r \sin(\phi)}{\sqrt{ r_{1,\parallel}^2 +  r_{1,\bot}^2} + \sqrt{\left(r_{1,\parallel} + r_{\parallel}\right)^2 + \left(r_{1,\bot} + r_{\bot}\right)^2 }} \right) 
\end{equation}

\begin{equation}
\begin{split}
    \alpha_{\mathrm{midpoint}} &= \arcsin\left(\frac{r \sin(\phi)}{2\sqrt{ r_{1,\parallel}^2 +  r_{1,\bot}^2}} \right) \\
    &+ \arcsin\left(\frac{r \sin(\phi)}{2\sqrt{\left(r_{1,\parallel} + r_{\parallel}\right)^2 + \left(r_{1,\bot} + r_{\bot}\right)^2 }} \right) \\
\end{split}
\end{equation}

\begin{equation}
    \alpha_{\mathrm{endpoint}} = \arctan\left(\frac{r_{\bot}}{\sqrt{ r_{1,\parallel}^2 +  r_{1,\bot}^2} + r_{\bot}}\right)
\end{equation}

The appendix~\ref{appendix:waimpact} is also showing the impact of the $\mathbf{d}$ vector definition on the contraction. For the plane-parallel case, which does not depend on the $\alpha$ angle, there is no need to provide a reference, and the definition of the line-of-sight reference does not change the result. Once the link between the basis is explicitly expressed, we can compute the contraction by computing the field covariance in the same way detailed in section~\ref{subsec:covariance}, but for a fixed grid of coordinates.

\subsection{Comparing covariance models}
\label{subsec:contraction_comparison}

Figure~\ref{fig:contraction} shows the contraction for the models of table~\ref{tab:models} on $(r_{\bot}, r_{\parallel})$ basis, considering as an illustration a reference $r_{1,\bot} = 0\ \mpc$ and $r_{1,\parallel} = 20\ \mpc$ to highlight the effect of the wide-angle model. To perform the covariance matrix calculation, the three power spectrum terms $P_{\mathrm{mm}}$, $P_{\mathrm{m\theta}}$, $P_{\mathrm{\theta\theta}}$ are computed using the modeling from~\cite{bel_accurate_2019} and the \texttt{CLASS} Boltzmann solver with nonlinear halofit correction. The wavenumbers used are logarithmically spaced in the range $k \in [0.0005, 1.0]\ h.\mathrm{Mpc}^{-1}$. We chose a rather large value of maximal wavenumber because this article aims to test our model and the previous ones and to extend their applicability to larger wavenumber ranges. The cosmological parameters used are following \texttt{AbacusSummit}~\citep{Maksimova2021,Garrison2021} baseline simulation. We also apply a grid window correction to the power spectrum for the simplest grid model (NGP), i.e. multiplying $P_{\mathrm{mm}}$ by $\Gamma(k, L)^2$ from equation~\ref{eq:grid_corr} ($n=1$), and $P_{\mathrm{m\theta}}$ by $\Gamma(k, L)$. We take a grid size value of $L = 10\ \mpc$, which is smaller than typical values used but it minimizes the impact of this mesh assignment correction. To compute the contraction, we express the $\Ai$ parameters of all models with the example values $f\sigma_8 = 0.4$, $b\sigma_8 = 0.8$, and $\beta_f = f/b = 0.5$. We use an RSD FoG parameter $\sigma_{\mathrm{g}} = 1.0\ \mpc$ and for the velocity side, we choose a value of $\sigma_{\mathrm{u}} = 15\ \mpc$, which is a commonly used value for this nuisance parameter~\citep{koda_are_2014,carreres_growth-rate_2023}.

For all $gg$ contractions, we see the effect of baryon acoustic oscillations (BAO) at $r \sim 100\ \mpc$ contained in the input power spectrum. The \textbf{AB17} model is expressed in a wide-angle framework but does not contain the RSD modeling in the $gg$ correlations, giving a fully isotropic correlation function. The \textbf{AB20} model contains RSD modeling without the wide-angle definition; thus, the $gv$ and $vv$ contractions are simpler (it is symmetric in $r_{\mathrm{\parallel}}$ but insufficient to model low-redshift field covariance. Finally, for the value of $\sigma_{\mathrm{g}}$ used in the contraction, the \textbf{L22} model and our \textbf{RC25} model are qualitatively and quantitatively equivalent. They accurately describe the RSD term in $gg$ while fully modeling the wide-angle for all the considered contractions. We verified that fixing the $\mathbf{r_1}$  reference to a very large value (e.g., $r_1 \sim 10$ Gpc) gives equal wide-angle and plane-parallel models with equivalent power spectrum models (e.g., \textbf{L22} and \textbf{RC25} compared to \textbf{AB20}). This means that our wide-angle models are correctly expressed within the limit of parallel lines-of-sight.

To estimate the validity range of \textbf{L22} and \textbf{RC25} models, we computed their contractions for different values of $\sigma_{\mathrm{g}}$ and report illustrative cases for one-dimensional $C_{gg}$ and $C_{gv}$ contractions in figure~\ref{fig:lai_instabilities}. The $\sigma_{\mathrm{g}}$ parameter is varied by a $1.0\ \mpc$ step. The $\sigma_{\mathrm{g}}$ value shown in the figure are the same as in figure~\ref{fig:contraction} (i.e. $1.0\ \mpc$), and the first values for which the two models start to disagree ($3.0\ \mpc$ for $gg$ and $7.0\ \mpc$ for $gv$). Our model is able to cover a larger $\sigma_{\mathrm{g}}$ range when integrating in the wavenumber range $k \in [0.0005, 1.0]\ h.\mathrm{Mpc}^{-1}$. For too large values of $\sigma_{\mathrm{g}}$, the \textbf{L22} model shows spurious oscillations that tend to increase in amplitude. The $gg$ contraction is the worst case and gives instabilities for $\sigma_{\mathrm{g}}$ values as low as $3.0\ \mpc$. We note that those instabilities are more important for radial separations $r_{\parallel}$. As in~\cite{lai_using_2022}, we verified that reducing the maximal wavenumber to $k_{\mathrm{max}} = 0.2\ h.\mathrm{Mpc}^{-1}$ improves the stability of \textbf{L22} model with respect to the $\sigma_{\mathrm{g}}$ parameter. This behavior from the  \textbf{L22} is expected since it is based on the assumption that $k \mu \sigma_{\mathrm{g}}$ is small compared to unity, i.e., reducing $k$ or $\mu$ stabilizes the model. 

Figure~\ref{fig:lai_instabilities2} shows the one-dimensional field covariance contraction for the galaxy-galaxy correlation for different maximal wavenumbers. For low maximal wavenumber $k_{\mathrm{max}} = 0.2\ h.\mathrm{Mpc}^{-1}$, since the galaxy-galaxy input power spectrum is still high at this typical wavenumber value, the cut-off of wavenumber is creating some unwanted oscillation pattern due to Hankel transform. This pattern is absent for $vv$ and $vg$ for which the input power spectrum is damped at large wavenumber. It indicates that using Hankel transform on an undamped power spectrum cut at a wavenumber that is too low is not advised. For higher maximal wavenumber $k_{\mathrm{max}} = 1.0\ h.\mathrm{Mpc}^{-1}$, the \textbf{L22} and \textbf{RC25} models give the same $gg$ contraction that is stable for both models. However, reaching $k_{\mathrm{max}} = 3.0\ h.\mathrm{Mpc}^{-1}$, \textbf{L22} becomes unstable due to the approximation used to derive this model, and these instabilities increase when $k_{\mathrm{max}}$ increases. Our model is stable with respect to maximal wavenumber and gives the same results for all values $k_{\mathrm{max}} > 1.0\ h.\mathrm{Mpc}^{-1}$ and for all field covariance contractions.

We conclude that our model is more stable for a wider range of wavenumber, orientation, and $\sigma_{\mathrm{g}}$ nuisance parameter. Furthermore, our model does not need to introduce an additional nuisance bias parameter $b_{\mathrm{add}}$ to integrate the small scales as is done in~\cite{adams_joint_2020} and~\cite{lai_using_2022}.

\subsection{Validation on N-body simulation}

\begin{figure*}
    \centering
	\includegraphics[width=0.9\textwidth]{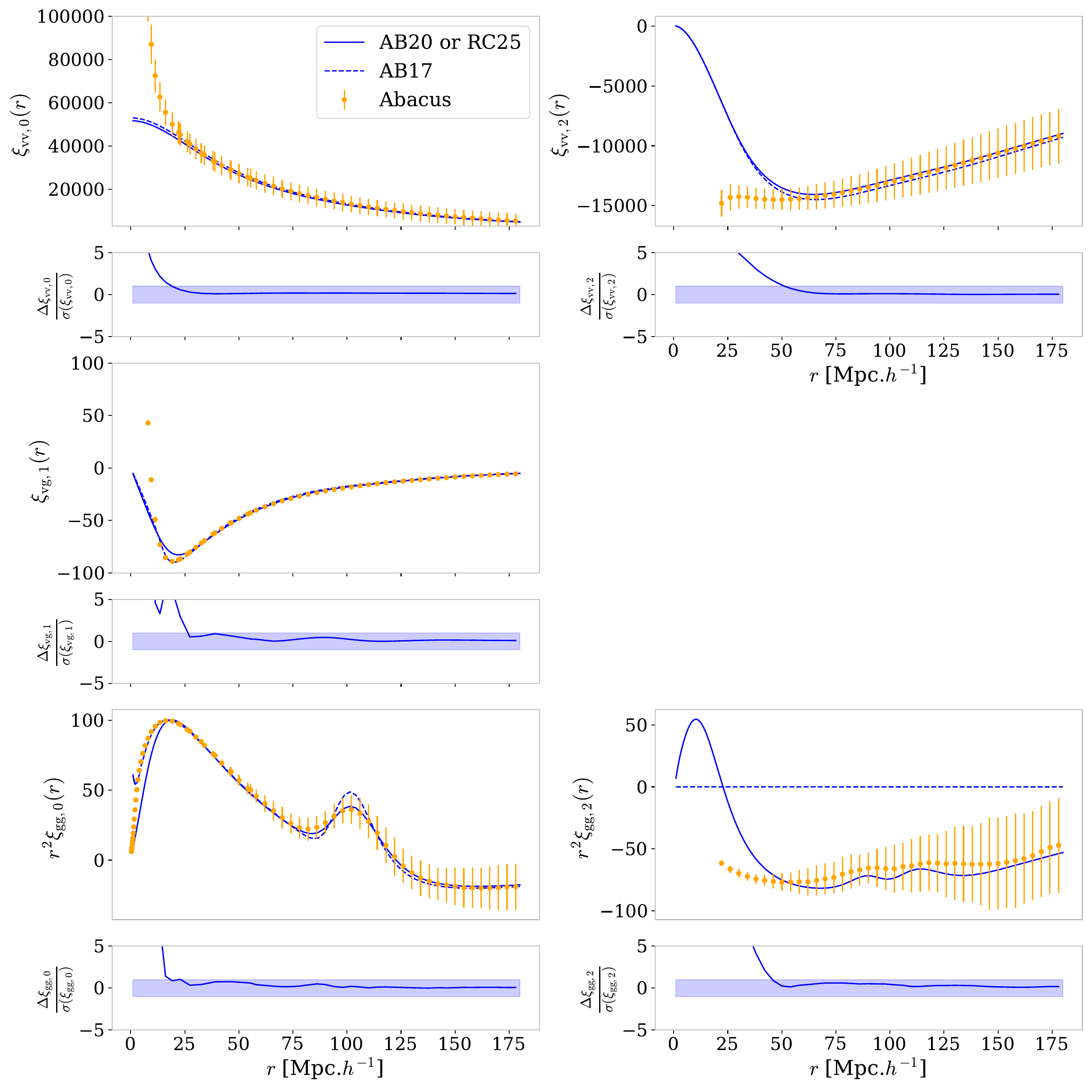}
    \caption{Representation of the $vv$ monopole (top left), quadrupole (top right), $vg$ dip (middle left), $gg$ monopole (bottom left) and quadrupole (bottom right) of the two-point correlation functions estimated on the \texttt{AbacusSummit} baseline halo catalog. The orange points are the average of the measurement over 25 simulations, and their associated error bar corresponds the standard deviation over those simulations, rescaled in volume to a sphere of maximal redshift $z=0.1$. The plain (\textbf{RC25}) and dashed (\textbf{AB17}) blue lines are the best-fit models determined with \texttt{flip} covariance contraction. The lower panel of each monopole shows the absolute difference between \textbf{RC25} model and the \texttt{AbacusSummit} multipoles, normalized by the multipole error bars.}
    \label{fig:abacus_contraction}
\end{figure*}

We want to show whether our new model can reproduce the clustering of an N-body simulation and at which scale it breaks down. It will notably give insight into the minimal separation that can be used with the current models. This study does not aim to develop a fitter for two-point correlation functions, as the \texttt{flip} formalism is not adapted for this, but rather to validate our field covariance models on a more compact representation.

We use the publicly available \texttt{AbacusSummit} N-body simulation \citep{Maksimova2021,Garrison2021}, and in particular, the halo catalogs of the 25 baseline $\Lambda$CDM cosmo000 (Planck 2018 cosmology) simulations. We assign galaxies to the dark matter halos of the simulations by using the halo occupation distribution (HOD) framework introduced in~\cite{Zheng2007}. We use a vanilla HOD with five parameters, which assumes that the number of galaxies $N$ in a given dark matter halo follows a probabilistic distribution that only depends on the halo's mass $M$. Moreover, the occupation is decomposed into the contribution of central and satellite galaxies  $\langle N(M) \rangle = \langle N_\textrm{cen}(M) \rangle + \langle N_\textrm{sat}(M) \rangle$. The number of central galaxy occupations follows a Bernoulli distribution, while the satellite galaxy occupation follows a Poisson distribution. In particular, we use the \texttt{AbacusHOD} implementation~\citep{Yuan2021}, available in the publicly available \texttt{abacusutils} \git{abacusorg/abacusutils}{} package. We chose the HOD five parameter ranges following the DESI Bright Galaxy Survey~\citep{Prada2023}, a typical sample for this kind of study. To reduce shot noise in our clustering measurements, we draw five independent realizations of galaxies over each halo catalog and using the same HOD model.

This study aims to measure the true clustering of velocity and density in the cosmic web. Therefore, we place ourselves in a flat-sky approximation, do not consider observational effects, and use the true velocities and densities of the galaxy catalog. For the density correlations, because we use a periodic box, we can make use of the simple Peebles estimator~\citep{peebles1974}:

\begin{equation}
\xi_{gg}(\mathbf{r})=\frac{D D(\mathbf{r})}{RR(\mathbf{r})}-1\, ,
\end{equation}

\noindent where $DD(\mathbf{r}) = \left\langle n_{\mathrm{g}}(\mathbf{r_1}) n_{\mathrm{g}}(\mathbf{r_2}) \right\rangle$ and $RR = \left\langle n_{\mathrm{r}}(\mathbf{r_1}) n_{\mathrm{r}}(\mathbf{r_2}) \right\rangle$ the normalized auto pair counts of the data and random catalogs. We consider the full galaxy catalog for computing velocity correlations, which means that we can use the same random catalog. Estimators for velocity autocorrelation and cross-correlation with galaxy density can be derived in a similar way as the Peebles estimator by defining the normalized auto galaxy momentum correlation $VV(\mathbf{r}) = \left\langle p_{\mathrm{g}}(\mathbf{r_1}) p_{\mathrm{g}}(\mathbf{r_2}) \right\rangle$ and cross momentum-density correlation $VD(\mathbf{r}) = \left\langle p_{\mathrm{g}}(\mathbf{r_1}) n_{\mathrm{g}}(\mathbf{r_2}) \right\rangle$. In the flat-sky approximation, the Peebles analog for peculiar velocities is given by

\begin{equation}
\xi_{vv}(\mathbf{r})=\frac{V V(\mathbf{r})}{RR(\mathbf{r})}, \quad \xi_{v g}(\mathbf{r})=\frac{V D(\mathbf{r})}{RR(\mathbf{r})}\, .
\end{equation}

The estimation of two-point correlation functions was performed using the \texttt{vegacorr} \git{TyannDB/vegacorr}{} Python package. Additionally, $\xi_{gg}$, $\xi_{gv}$, and $\xi_{vv}$ were decomposed into multipoles $\xi_{ab,\ell}$ following equation~\ref{eq:multipoles}. We kept the monopole ($\ell=0$) and quadrupole ($\ell=2$) of the $vv$ and $gg$ terms, and the dipole ($\ell = 1$) of the $vg$ term, as they are the only non-null terms. We computed the error bar associated with each multipole as the standard deviation over the 25 realizations. Given the large volume of the simulations used for two-point correlation estimation ($8\ \mathrm{Gpc}^3.h^{-3}$), we scale uncertainties up to match smaller volumes typically measured in peculiar velocity surveys, i.e. $1.12\ \mathrm{Gpc}^3.h^{-3}$ for a sphere up to $z < 0.1$.

Figure~\ref{fig:abacus_contraction} shows the multipoles computed on the \texttt{AbacusSummit} simulation, as well as the best-fit \texttt{flip} contracted model. Those models are generated similarly to section~\ref{subsec:contraction_comparison} but without the grid window correction since gridding is not used for building the two-point correlation functions. We set $k_{\mathrm{max}} = 100 h.\mathrm{Mpc}^{-1}$ to avoid aliasing in the integration without the grid window function. The fit was performed in contracted space directly with \texttt{iminuit} by varying the parameters $\beta_f$, $f\sigma_8$, $b\sigma_8$, $\sigma_{\mathrm{u}}$, and $\sigma_{\mathrm{g}}$. Since the errors are extremely underestimated at small scales, we used in the fitting procedure a constant uncertainty for each multipole $\xi_{ab,\ell}$ equal to the maximal error bar. The $\chi^2$ function used in the fit is then the sum of $\chi^2$ for each multipole. Since our model is not able to reproduce the very small scales of the two-point correlation functions, very sensitive to the HOD model considered, we take a minimal fitting value of $r_{\mathrm{min}} = 20\ \mpc$ for the $\ell=0$ and $\ell= 1$ multipoles, and $r_{\mathrm{min}} = 40\ \mpc$ for the quadrupoles. To match our measurements, we do not include wide-angle effects in this fit. Consequently, we use our model \textbf{RC25} expressed at large reference separation, equivalent to the \textbf{AB20} model.

For comparison, we also show the best-fit \textbf{AB17} model, which does not model RSD in the $gg$ and $gv$ correlations. We clearly see that \textbf{AB17} fails to model the $gg$ quadrupole (by construction) as well as the damping of the BAO peak caused by the FoG RSD effect. 

The \textbf{RC25} model gives a reasonable agreement with the N-body simulation for scales $r \gtrapprox 15\ \mpc$ for monopoles ($\xi_{vv,0}$ and $\xi_{gg,0}$), $r \gtrapprox 25\ \mpc$ for $\xi_{vg,1}$, and $r \gtrapprox 45\ \mpc$ for quadrupoles ($\xi_{vv,2}$ and $\xi_{gg,2}$). This model is suitable for large linear scales but should be taken with care in the intermediate range scale $15 < r < 45\ \mpc$ and ultimately breaks down for the smallest scales. Those discrepancies highlight the potential room for improvement of our model. In particular, we expect nonlinear extensions of the model to improve small scales. On the $gg$ correlation, another potential improvement would be to include HOD modeling. On the $vv$ side, both monopole and quadrupole are controlled only by two parameters ($f\sigma_8$ and $\sigma_{\mathrm{u}}$), which does not seem to be sufficient for modeling all scales. Our study exposes the need for better nonlinear models of velocity clustering, such as~\cite{dam_exploring_2021}. Finally, we note that this fit is not optimal. We can improve it with a full covariance matrix treatment and better data uncertainties accounting for HOD variability, but this is out of the scope of this study. Furthermore, this analysis represents an extremely accurate noise-free measurement, far beyond current or future observational surveys. The suggested enhancements in the modeling may not be needed when treating current datasets.

\section{Survey-dependent Fisher forecast}
\label{sec:fisher}

\begin{figure}
	\includegraphics[width=0.95\columnwidth]{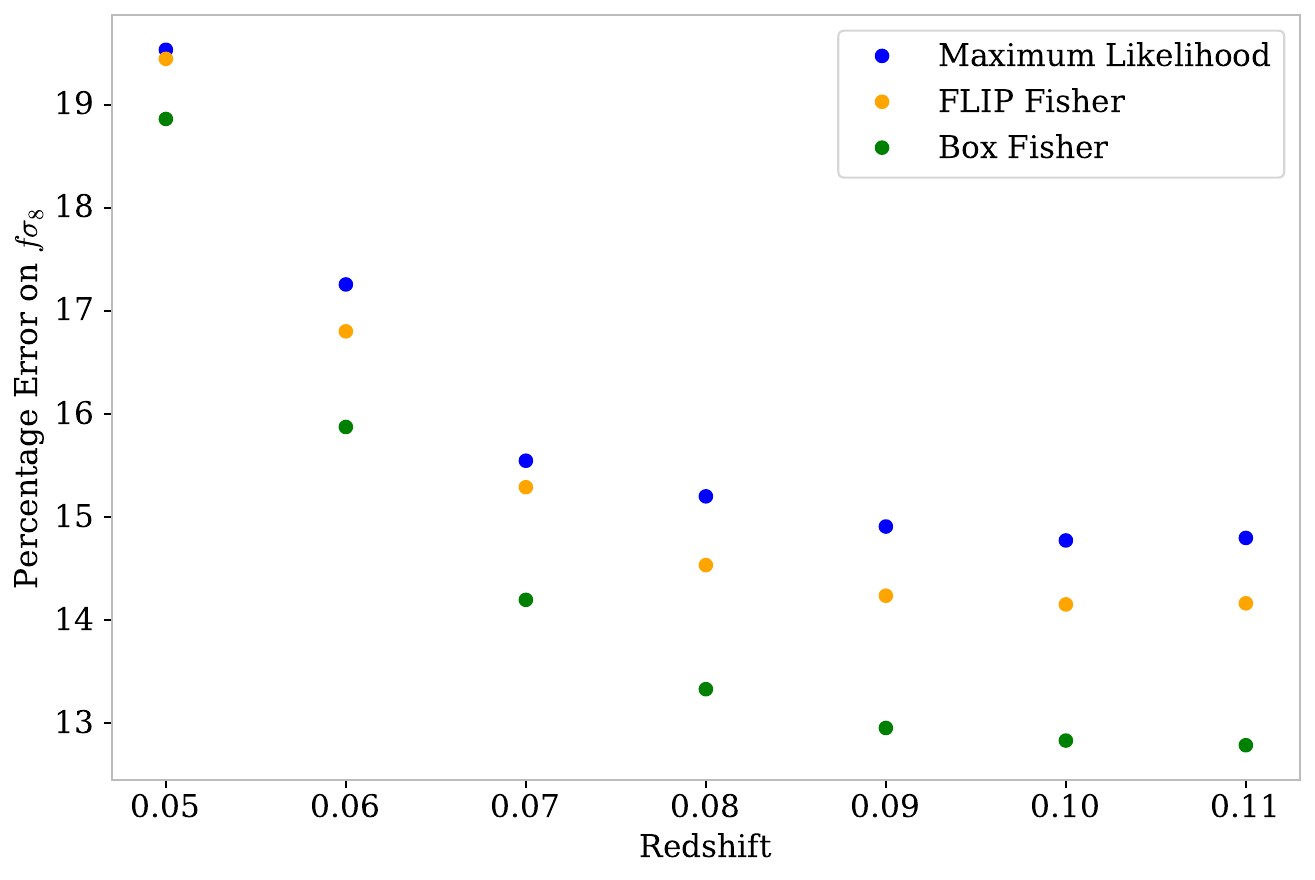}
    \caption{Percentage error on $f\sigma_8$ for different redshift bins. The orange points show the results for our \texttt{flip} Fisher forecast, the green points the results for the standard volume Fisher, and the blue points the ones for the likelihood-based method. The results are obtained for different redshift bins with the same minimum redshift, $z_{min}=0.02$, and increasing maximum redshift. The results are plotted at the maximum redshift of each bin and are the average over 27 realizations of a ZTF 6-year survey.}
    \label{fig:ML_vs_fisher}
\end{figure}

The framework we detail in section~\ref{sec:covariance} allows for the fast generation of field covariance matrices following the geometry of a given data sample. The likelihood-based method is time-consuming mainly due to the inversion of the covariance matrix at each likelihood calculation, which can reach many occurrences for minimization or MCMC sampling. \texttt{Flip} implements a faster way to estimate the uncertainties of parameters of interest by computing the Fisher information from the field covariance.
 
A standard Fisher forecast~\citep{koda_are_2014,howlett_measuring_2017} is based on the information contained in the peculiar velocity power spectrum. In this method, the Fisher information matrix is computed using

\begin{equation}
\begin{aligned}
& F_{i j}=\frac{\Omega_{\mathrm{s k y}}}{4 \pi^2} \int_{r_{\min }}^{r_{\max }} r^2 d r \int_{k_{\min }}^{k_{\max }} k^2 d k \int_0^1 d \mu \operatorname{Tr}\left[\mathbf{C^H}^{-1} \mathbf{C^H_{,i}} \mathbf{C^H}^{-1}   \mathbf{C^H_{,j}} \right]\, ,
\end{aligned} 
\end{equation}

\noindent where $\Omega_{\mathrm{s k y}}$ is the solid angle from the sky coverage of the survey, $r_{\min}$ and $r_{\max}$ are the lower and upper limits of the considered survey in term of comoving distance, $k_{\min}$ and $k_{\max}$ defines the wavenumber of integration. The covariance matrix $\mathbf{C^H}$ is defined in Fourier space for velocity-only Fisher forecast as

\begin{equation}
    \mathbf{C^H}=\left[\begin{array}{cc} P_{v v}\left(k, \mu\right)+\frac{\sigma_{obs}^2(r)}{\bar{n}_v(r)}\, ,
\end{array}\right]
\end{equation}

\noindent where a shot-noise term is added with the typical peculiar velocity uncertainty $\sigma_{obs}(r)$, and the number density for the considered velocity tracer $\bar{n}_v(r)$. The positions of each velocity tracer is not taken into account. Only the redshift distribution of the considered sample and the survey area probed are considered. We created a Python version \git{DamianoRosselli/fisher_howlett}{} of the code given in~\cite{howlett_measuring_2017}, and use it to derive the expected uncertainties on $f\sigma_8$.

In contrast, our Fisher forecast is computed from the precise position of each data point in the considered field and can account for geometrical effects and velocity/density distribution in a specific survey configuration. Additionally, our forecast can account for various systematic effects thanks to the error vector construction and the potential addition of an observational field covariance matrix. We computed the Fisher information matrix following~\cite{Tegmark1996}, in the Gaussian case:

\begin{equation}\label{eq:fisher_matrix}
    F_{i j}=\langle\mathcal{L}_{, ij}\rangle=\frac{1}{2} \operatorname{Tr}\left[\boldsymbol{A}_i \boldsymbol{A}_j+\boldsymbol{C}^{-1} \boldsymbol{M}_{i j}\right]\, ,
\end{equation}

\noindent with the matrices defined by $\boldsymbol{A}_i = \boldsymbol{C}^{-1} \boldsymbol{C}_{,i}$ and $\boldsymbol{M}_{i j}=\boldsymbol{\mu}_{,i} \boldsymbol{\mu}_{,j}^t+\boldsymbol{\mu}_{,j} \boldsymbol{\mu}_{,i}^t$ where the $\boldsymbol{\mu}$ is the average of the considered field vector $\boldsymbol{x}$. This expression can be simplified in our case because we are considering only vectors with null averages (density, velocity, and logarithmic distance), i.e., we have $\boldsymbol{M}_{i j} = 0$.

In the formalism of \texttt{flip}, considering the parameters of interest $\Ai$, we can compute the covariance derivatives from equation \ref{eq:sum_cov} such that:

\begin{equation}
    \Cab_{,i} = \sum_n  \frac{\partial \Ai}{\partial \Aii} \Cabi\, .
\end{equation}

Using the covariance derivatives, we build the Fisher information matrix given in Eq. \eqref{eq:fisher_matrix}, and by inverting the Fisher matrix, we obtain the expected uncertainties on the parameters of interest.

We test our Fisher forecast formalism on a ZTF-like simulation of SN~Ia peculiar velocities. Specifically, we analyze 27 realizations of a 6-year ZTF survey simulated by \cite{carreres_growth-rate_2023} using the \texttt{snsim} \git{bastiencarreres/snsim}{} package. We use the peculiar velocities extracted from the ZTF simulations to compare the estimated error bar on $f\sigma_8$ between \texttt{flip} Fisher, a standard volume Fisher, and a full likelihood-based study.

Figure \ref{fig:ML_vs_fisher} shows the results for the percentage error on $f\sigma_8$ of our \texttt{flip} Fisher forecast, compared to the standard volume Fisher and the likelihood-based method. The likelihood-based results are taken from figure 13 in \cite{carreres_growth-rate_2023}. Our Fisher method achieves errors that are more comparable to the likelihood-based analysis with respect to the standard volume Fisher forecast, especially for larger redshift ranges. We interpret the inaccuracy of the volume Fisher forecast to be caused by the data geometry, and the nuisance parameters not being taken into account. On average, over all the redshift bins, our Fisher method gives uncertainties $30$ \% closer to the likelihood-based estimate in comparison to the standard volume one.

\section{Conclusion and future prospects}
\label{sec:prospect}

In this work, we have developed a generalized framework for inferring the growth rate of large-scale structures with the likelihood-based field-level method. The generality of this framework also makes it work with a large variety of peculiar velocity probes (TF, FP, and \sn) and galaxy density estimators. We implemented most of previously used models in the literature and we introduced a new density and velocity covariance model that accounts for wide-angle effects, redshift-space distortions and FoG with better numerical convergence. The covariance generation, the likelihood-based inference, and the wide variety of vector estimators are all implemented consistently in the \texttt{flip} software. 

This generalized framework includes several applications that allow us to validate it. First, we compared the field covariance generation to a previous code publicly available. We derived inside our framework the field covariance models derived previously in the literature~\citep{adams_improving_2017,adams_joint_2020,lai_using_2022,carreres_growth-rate_2023} and compared them to our new model by expressing the covariance in regular gridding. This contraction of the covariance matrix allows us to test the validity of the models with respect to wavenumber, fitted parameters, and orientation ranges. In particular, we concluded that the model we developed in this work is the most suitable for covering a larger range of wavenumber, orientation, and RSD FoG nuisance parameters than previous models while keeping full wide-angle modeling. By comparing this contraction to two-point correlation functions measured on N-body simulations, we obtain insights into the minimal separation for which our model is still valid.

Secondly, as our framework allows for an efficient computation of field covariance, we developed a Fisher forecast method that accounts for the geometry of the considered survey. We validated this method on an N-body simulation by comparing its performance to a standard volume Fisher forecast software and a full likelihood-based inference. This comparison was performed on a ZTF-like simulation of \sn~peculiar velocities. We concluded that since our Fisher forecast method accounts for the entire geometry of the survey, i.e., the position of all the objects used, it gives error bars $30$ \% closer to the full likelihood-based estimate on average in comparison to the standard volume method.

We identify several avenues for extending this work. The most-straightforward improvement is the inclusion of nonlinear models such as the TNS model~\citep{Taruya2010}, the regularized PT model at two-loop order model~\citep{Taruya2012}, or the effective field theory of large scale structures (EFTofLSS, see e.g.,~\cite{Carrasco2013,Perko2016,DAmico2020,Ivanov2022}) in the likelihood-based framework. Such models will increase the accuracy of fits on small scales, enabling smaller mesh cell sizes to be used for the density field. The latter can help reducing the uncertainty on $f\sigma_8$.

The generalized field covariance framework we developed can also be extended to the covariance of two-point compressed statistics such as a two-point correlation function or power spectrum, allowing one to integrate complex power spectrum models. It can be done by adapting the equations presented in \cite{blake_correlations_2023} to our generalized framework.

The likelihoods and vector estimators developed in our framework can be improved. In particular, we can develop more complex non-Gaussian likelihoods to include observational effects such as selection effects due to magnitude cuts of the considered tracer. Our framework was mainly built to deal with SN~Ia peculiar velocities, particularly to simultaneously fit Hubble diagram parameters along with cosmological ones. We plan to adapt this simultaneous fitting to the case of FP and TF peculiar velocity estimators. Those estimators contain normalization parameters characteristic of both empirical relations, and adapting our method will allow us to account for all parameter correlations. 

Our survey geometry-dependent Fisher forecast method and the likelihood-based method can be applied to a larger variety of surveys. With the Fisher forecast, we plan to forecast the cosmic variance for the geometry of recent and future surveys, i.e., for which number of data densities the improvement on the $f\sigma_8$ error no longer evolves. On the likelihood-based method side, we plan to apply it to the Zwicky Transient Facility (ZTF)~\citep{ztf2019,Rigault2024} and the Legacy Survey of Space and Time Rubin-(LSST)~\citep{ivezic_lsst_2018} for \sn. In addition, the \texttt{flip} software will allow us to combine SN~Ia surveys with galaxy surveys such as the Dark Energy Spectroscopic Instrument (DESI)~\citep{desi_collaboration_desi_2016,desi_collaboration_desi_2016-1,martini_overview_2018} with the peculiar velocity studies~\citep{Saulder2023,Said2024} or the 4 meter Multi-Object Spectroscopic Telescope (4MOST)~\citep{de_jong_4most_2019} with the combination of the Cosmology Redshift Survey (4MOST-CRS)~\citep{richard_4most_2019} with the 4MOST Hemisphere Survey of the Nearby Universe (4MOST-4HS)~\citep{taylor_4most_2023}. Finally, the framework we developed will be necessary for consistently combining the fit of all the different samples of peculiar velocities. We plan to extend our framework so that it will be able to treat simultaneously SN~Ia peculiar velocities (ZTF, LSST), FP/TF peculiar velocities from galaxy surveys (DESI, 4MOST-4HS), and the galaxy density field (DESI, 4MOST-CRS, 4MOST-4HST).

\begin{acknowledgements}

The project leading to this publication has received funding from Excellence Initiative of Aix-Marseille University - A*MIDEX, a French ``Investissements d'Avenir'' program (AMX-20-CE-02 - DARKUNI). 
AGK was supported by the U.S. Department of Energy (DOE), Office
of Science, Office of High-Energy Physics, under Contract No. DE–AC02–05CH11231.

\end{acknowledgements}

\bibliographystyle{aa}
\bibliography{bibli_pv}

\begin{appendix}

\section{Spherical harmonic properties}

\subsection{Spherical coordinates}
It is useful to integrate over $\mathbf{k}$ using the spherical coordinates as
\begin{align}
    \label{eq:spherical_decomposition}
    \int_{\mathbf{k}} d^3 \mathbf{k} &= \intsphere \\ 
    &= \int_k k^2 dk\int_\Omega d\Omega
\end{align}

\subsection{Legendre polynomial expansion}
In the \texttt{flip} formalism we take advantage of the Legendre polynomial expansion. This expansion gives that a function $f(x)$ with $x\in [-1, 1]$ can be expanded as

\begin{equation}
    \label{eq:LegendreExp}
    f(x) = \sum_{l=0}^{+\infty} a_\ell L_\ell(x)\, ,
\end{equation}

\noindent with the coefficient $a_\ell$ such that

\begin{equation}
    \label{eq:multipoles}
    a_\ell = \frac{2\ell + 1}{2}\int_{-1}^1 dx^\prime f(x^\prime) L_\ell(x^\prime)\, .
\end{equation}

One useful Legendre expansion is the plane-wave expansion:

\begin{equation}
    \label{eq:planewave}
    e^{i\mathbf{k}\cdot\mathbf{r}} = \sum_{l=0}^{+\infty} (2\ell + 1) i^\ell j_\ell(kr)L_\ell(\mathbf{\hat{k}}\cdot\mathbf{\hat{r}})\, .
\end{equation}

\subsection{Product of Legendre polynomials}

A useful theorem to express the product of Legendre polynomials is the spherical harmonic addition theorem, which gives

\begin{equation}
    \label{eq:spherical_harmonic}
    L_{\ell}(\hat{\mathbf{k}} . \hat{\mathbf{r}})=\frac{4 \pi}{(2 \ell+1)} \sum_{m=-\ell}^{\ell} Y_{\ell m}(\hat{\mathbf{k}}) Y_{\ell m}(\hat{\mathbf{r}})^* \, .
\end{equation}

\subsubsection{Product of three Legendre polynomials}

From Eq.~\ref{eq:spherical_harmonic} we obtains that

\begin{equation}
    \begin{split}
            &\Lg{}(\normv{k}\cdot\normv{r})\Lg{1}(\normv{k}\cdot\normv{r}_1)\Lg{2}(\normv{k}\cdot\normv{r}_2) = \frac{(4\pi)^3}{(2 \ell+1)(2 \ell_1+1)(2 \ell_2+1)}\\
            &\times\sum_{m=-\ell}^{\ell}\sum_{m_1=-\ell_1}^{\ell_1}\sum_{m=-\ell_2}^{\ell_2}\Ylm{}{}(\normv{k})\Ylm{1}{}(\normv{k})\Ylm{2}{}(\normv{k})\\
            &\times
            \Ylms{}{}(\normv{r})\Ylms{1}{}(\normv{r}_1)\Ylms{2}{}(\normv{r}_2)\, .
    \end{split}
\end{equation}

The Gaunt coefficients are defined such that

\begin{equation}
    \begin{split}
    \Gaunt&=\int_{\Omega}d\Omega \Ylm{}{}(\normv{k})\Ylm{1}{}(\normv{k})\Ylm{2}{}(\normv{k})\\
    &=\sqrt{\frac{(2 \ell+1)(2 \ell_1+1)(2 \ell_2+1)}{4\pi}}\\
    &\times
    \begin{pmatrix}
        \ell & \ell_1 & \ell_2 \\
                     0   &   0    &   0    \\
    \end{pmatrix}
        \begin{pmatrix}
        \ell & \ell_1 & \ell_2 \\
        m   &   m_1    &   m_2    \\
    \end{pmatrix}\, ,
    \end{split}
\end{equation}

\noindent where we used the Wigner 3-$j$ symbols.

We then can express the integral of the product of three Legendre polynomials over a solid angle as
\begin{equation}\label{eq:intLLL}
    \begin{split}
     &\int_\Omega d\Omega\Lg{}(\normv{k}\cdot\normv{r})\Lg{1}(\normv{k}\cdot\normv{r}_1)\Lg{2}(\normv{k}\cdot\normv{r}_2)
     = \frac{(4\pi)^3}{(2 \ell+1)(2 \ell_1+1)(2 \ell_2+1)}\\
        &\times\sum_{m=-\ell}^{\ell}\sum_{m_1=-\ell_1}^{\ell_1}\sum_{m=-\ell_2}^{\ell_2}\Gaunt\Ylms{}{}(\normv{r})\Ylms{1}{}(\normv{r}_1)\Ylms{2}{}(\normv{r}_2)\, .
    \end{split}
\end{equation}

\subsubsection{Product of two Legendre polynomials}

From Eq.~\ref{eq:spherical_harmonic} we obtains that
\begin{equation}
    \begin{split}    
    &\Lg{}(\normv{k}\cdot\normv{r})L_{\ell^\prime}(\normv{k}\cdot\normv{d}) = \frac{(4\pi)^2}{(2\ell + 1)(2\ell^\prime + 1)}\\
    &\times\sum_{m=-\ell}^\ell \sum_{m^\prime=-\ell^\prime}^{\ell^\prime} \Ylm{}{}(\normv{k})\Ylms{}{}(\normv{r})\Ylms{}{\prime}(\normv{k})\Ylm{}{\prime}(\normv{d})
    \end{split}\, .
\end{equation}

Spherical harmonics are orthonormal, this property can be expressed as

\begin{equation}
    \int_{\Omega} d\Omega \Ylm{}{}\Ylms{}{\prime} = \delta^K_{\ell\ell^\prime}\delta^K_{mm^\prime}\, ,
\end{equation}
where $\delta^K$ is the Kronecker's delta.

We then obtain
\begin{equation}\label{eq:int2L}
        \int_{\Omega} d\Omega\Lg{}(\normv{k}\cdot\normv{r})L_{\ell^\prime}(\normv{k}\cdot\normv{d})=\frac{(4\pi)^2\delta^K_{\ell\ell^\prime}}{(2\ell + 1)^2}\sum_{m=-\ell}^\ell\Ylm{}{}(\normv{d})\Ylms{}{}(\normv{r})\, .
\end{equation}

\section{Extension of covariance calculation with redshift dependency}
\label{appendix:redshift_dep}

Both wide-angle and plane-parallel power spectrum models can be extended to directly account for the redshift dependency of the parameters to fit. We consider that this redshift evolution can be written at the power spectrum stage such that:

\begin{equation}
\Ai = \Aai(z_1) \Abi(z_2)\, ,
\end{equation}

\noindent where $z_1$ and $z_2$ are the redshifts of the considered pair. This decomposition is always valid when considering linear models for velocities and densities, both with and without wide-angle.

The covariance of one pair of points can then be written:

\begin{equation}
    \Cab = \sum_n  \Aai(z_1) \Abi(z_2) \Cabi\, ,
\end{equation}

\noindent and the full covariance can be easily computed using the outer product $\Aai$ and $\Abi$ for all redshifts considered.

The redshift dependencies are optionally implemented in the \texttt{flip} software. This extension is used to account for the redshift dependence of the growth rate in velocity terms, often parameterized with the growth index $\gamma$ such that

\begin{equation}
\Av(z) = f(z) \sigma_8(z) =  \Omega_m(z)^{\gamma} \sigma_8(z)\, .
\end{equation}

\section{Vector field covariance in the \texttt{flip} framework}
\label{appendix:vector_cov}

Let's consider a vector field $\mathbf{x}$ in the basis ${x, y, z}$, such as that for the velocity vector field. The vector correlation tensor is defined by 

\begin{equation}
     \langle \mathbf{x_{1, i} x_{2, j} }\rangle = \Psi_{ij}(r) = 
            \Psi_\perp(r) \delta_{ij} + \left[\Psi_\parallel(r) - \Psi_\perp(r)\right] r_i r_j\, ,
\end{equation}

\noindent where the indexes $i$ and $j$ vary for all axes ${x, y, z}$. By projecting the $(\mathbf{r_{\perp}}, \mathbf{r_{\parallel}})$ basis into ${x, y, z}$ we can obtain the following expression for the tensor components:

\begin{align}
    \Psi_\perp(r) &= \int_0^{+\infty} \frac{1}{3}\left[j_0(kr) + j_2(kr)\right]P_{xx}(k)dk \, , \\
    \Psi_\parallel(r) &= \int_0^{+\infty} \frac{1}{3}\left[j_0(kr) - 2j_2(kr)\right]P_{xx}(k)dk \, .  
\end{align}

For any directional unit vector $\mathbf{\hat{n}}_1$ and $\mathbf{\hat{n}}_2$ the correlation of the projected vector field is then defined by:

\begin{align}
    \langle (\mathbf{x}_{1}\cdot\mathbf{\hat{n}_1})(\mathbf{x}_{2}\cdot\mathbf{\hat{n}_2}) \rangle =& \Psi_\perp(r) (\mathbf{\hat{n}_1}\cdot\mathbf{\hat{n}_2}) \\
    &+ \left[\Psi_\parallel(r) - \Psi_\perp(r)\right](\mathbf{\hat{r}}_{1}\cdot\mathbf{\hat{n}_1})(\mathbf{\hat{r}}_{2}\cdot\mathbf{\hat{n}_2})\, .
\end{align}

Similarly to the formalism developed in section~\ref{sec:covariance}, we can express the field covariance matrix of projected vectors by performing additional Legendre expansions for the directional vectors. An expression for the generalized field covariance of two vector fields $\mathbf{a}$ and $\mathbf{b}$ projected over the direction $\mathbf{\hat{n}}_1$ and $\mathbf{\hat{n}}_2$ is then given by:

\begin{equation}
    \begin{split}
        &C_{(\mathbf{a}\cdot\mathbf{\hat{n}_1})(\mathbf{b}\cdot\mathbf{\hat{n}_2}), n}(\mathbf{r_1}, \mathbf{r_2}) = \sum_l i^l \int_k \frac{k^2 dk}{2 \pi^2} P_{ab}(k) j_l(k) \\
        &\times (4\pi)^4  \sum_{l_1, l_2,l_3, l_4}\sum_{m, m_i =-l, -l_i}^{+l, +l_i} Y_{lm}^*(\hat{r})Y^*_{l_1m_1}(\hat{n}_1)Y^*_{l_2m_2}(\hat{n}_2)Y^*_{l_3m_3}(\hat{r}_1)Y^*_{l_4m_4}(\hat{r}_2)\\
        &\times \sum_{c_1c_2} \sum_{\gamma_1, \gamma_2 = -c_1, -c_2}^{+c_1, +c_2} (-1)^{\gamma_1+\gamma_2} G_{m, \gamma_1, \gamma_2}^{l, c_1, c_2} G_{m_1, m_2, -\gamma_1}^{l_1, l_2, c_1}G_{m_3, m_4, -\gamma_2}^{l_3, l_4, c_2}\\
        &\times\frac{1}{16}\int_{-1}^1L_{l_1}(\xi_1)L_{l_2}(\xi_2)L_{l_3}(\mu_1)L_{l_4}(\mu_2)B(k, \xi_1, \xi_2, \mu_1, \mu_2) \\
        &\times d\xi_1d\xi_2d\mu_1d\mu_2\, ,
    \end{split}
\end{equation}

\noindent where $\boldsymbol{r} = \boldsymbol{r}_2 - \boldsymbol{r}_1$, $\xi_1 = \mathbf{\hat{k}}\cdot\mathbf{\hat{n}}_1$, $\xi_2 = \mathbf{\hat{k}}\cdot\mathbf{\hat{n}}_2$,  $\mu_1 = \mathbf{\hat{k}}\cdot\mathbf{\hat{r}}_1$ and  $\mu_2 = \mathbf{\hat{k}}\cdot\mathbf{\hat{r}}_2$. We make note that this expansion is not yet implemented in the \texttt{flip} framework.

\section{Decomposition of previous models in the \texttt{flip} framework}
\label{appendix:table_models}

The detail decomposition of \textbf{AB17}~\cite{adams_improving_2017}, \textbf{AB20}~\cite{adams_joint_2020}, \textbf{L22}~\citep{lai_using_2022}, and \textbf{RC25} (This work) models in the \texttt{flip} framework detailed in section~\ref{sec:covariance} is given in table~\ref{tab:models}

\begingroup

\renewcommand{\arraystretch}{1.4}

\begin{table*}
	\centering
    \caption{Decomposition, of all the power spectra models used in~\cite{adams_improving_2017,adams_joint_2020,lai_using_2022,carreres_growth-rate_2023}, and the one developed in this study.}
    \label{tab:models}
    \resizebox{0.99\textwidth}{!}{\begin{tabular}{ccccc}
    \hline\hline
    \multicolumn{5}{c}{\textbf{AB17} \citep{adams_improving_2017} - Wide-angle model without RSD on density} \\
    \hline\hline
     Fields $ab$ & Term number $n$ & $\Ai$ & $\Bi$ & $\Pabi$ \\
    \hline
    $gg$ & $0$ & $(b\sigma_8)^2$  & $1$ & $P_{\mathrm{mm}}(k)$ \\
    \hline
    $gv$ & $0$ & $b\sigma_8 f\sigma_8$  & $(iaH) \frac{\mu_2}{k} $ & $P_{\mathrm{m\theta}} (k) D_{\mathrm{u}}(k,\sigma_{\mathrm{u}})$ \\
    \hline
    $vv$ & $0$ & $(f\sigma_8)^2$  & $(aH)^{2} \frac{\mu_1 \mu_2}{k^2} $ & $P_{\mathrm{\theta\theta}} (k) D_{\mathrm{u}}^2(k,\sigma_{\mathrm{u}})$ \\
    \hline\hline
    \multicolumn{5}{c}{\textbf{AB20} \citep{adams_joint_2020} - Plane-parallel model with RSD on density} \\
    \hline\hline
     Field $ab$ & Term $n$ & $\Ai$ & $\Bi$ & $\Pabi$ \\
    \hline
    & $0$ & $(b\sigma_8)^2$  & $\exp\left[-(k\sigma_{\mathrm{g}}\mu)^2\right]$ & $P_{\mathrm{mm}}(k)$ \\
    $gg$ & $1$ & $(b\sigma_8)^2 \beta_f$  & $2 \mu^2 \exp\left[-(k\sigma_{\mathrm{g}}\mu)^2\right]$ & $P_{\mathrm{m\theta}}(k)$ \\
    & $2$ & $(b\sigma_8)^2 \beta_f^2$  & $\mu^4 \exp\left[-(k\sigma_{\mathrm{g}}\mu)^2\right]$ & $P_{\mathrm{\theta\theta}}(k)$ \\
    \hline
    $gv$ & $0$ & $b\sigma_8 f\sigma_8$  & $(iaH) \frac{\mu}{k} \exp\left[-\frac{(k \sigma_{\mathrm{g}}\mu)^2}{2}\right]$ & $P_{\mathrm{m\theta}} (k) D_{\mathrm{u}}(k,\sigma_{\mathrm{u}})$ \\
    & $1$ & $(f\sigma_8)^2$  & $(iaH)\frac{\mu^3}{k} \exp\left[-\frac{(k \sigma_{\mathrm{g}}\mu)^2}{2}\right]$ & $P_{\mathrm{\theta\theta}} (k) D_{\mathrm{u}}(k,\sigma_{\mathrm{u}})$ \\
    \hline
    $vv$ & $0$ & $(f\sigma_8)^2$  & $(aH)^{2} \frac{\mu^2}{k^2} $ & $P_{\mathrm{\theta\theta}} (k) D_{\mathrm{u}}^2(k,\sigma_{\mathrm{u}})$ \\
    \hline\hline
    \multicolumn{5}{c}{\textbf{L22} \citep{lai_using_2022} - Wide-angle model with RSD and Taylor expansion of FoG} \\
    \hline\hline
     Field $ab$ & Term $n$ & $\Ai$ & $\Bi$ & $\Pabi$ \\
    \hline
    & $0, m$ & $(b\sigma_8) ^2 \sigma_{\mathrm{g}}^{2m}$  & $\sum_{p,q, p+q = m} \left(\frac{(-1)^{p+q}}{2^{p+q} p! q!}\right) k^{2(p+q)} \mu_1^{2p} \mu_2^{2q}$ & $P_{\mathrm{mm}}(k)$ \\
    $gg$ & $1, m$ & $ (b\sigma_8)^2 \beta_f \sigma_{\mathrm{g}}^{2m}$  & $\sum_{p,q, p+q = m} \left(\frac{(-1)^{p+q}}{2^{p+q} p! q!}\right) k^{2(p+q)} \mu_1^{2p} \mu_2^{2q} (\mu_1^{2} + \mu_2^{2} )$ & $P_{\mathrm{m\theta}}(k)$ \\
    & $2, m$ & $(b\sigma_8)^2 \beta_f^2 \sigma_{\mathrm{g}}^{2m}$  & $\sum_{p,q, p+q = m} \left(\frac{(-1)^{p+q}}{2^{p+q} p! q!} \right)k^{2(p+q)} \mu_1^{2p+2} \mu_2^{2q+2}$ & $P_{\mathrm{\theta\theta}}(k)$ \\
    \hline
    $gv$ & $0, m$ & $(b\sigma_8)^2 \beta_f \sigma_{\mathrm{g}}^{2m}$  & $(iaH) \left(\frac{(-1)^m}{2^m m!}\right)  k^{2m-1} \mu_2 \mu_1^{2m}$ & $P_{\mathrm{m\theta}} (k) D_{\mathrm{u}}(k,\sigma_{\mathrm{u}})$ \\
    & $1, m$ & $(f\sigma_8)^2 \sigma_{\mathrm{g}}^{2m}$  & $(iaH) \left(\frac{(-1)^m}{2^m m!}\right) k^{2m-1} \mu_2 \mu_1^{2m+2}$ & $P_{\mathrm{\theta\theta}} (k) D_{\mathrm{u}}(k,\sigma_{\mathrm{u}})$ \\
    \hline
    $vv$ & $0$ & $(f\sigma_8)^2$  & $(aH)^{2} \frac{\mu_1 \mu_2}{k^2}$ & $P_{\mathrm{\theta\theta}} (k) D_{\mathrm{u}}^2(k,\sigma_{\mathrm{u}})$ \\
    \hline\hline
    \multicolumn{5}{c}{\textbf{RC25} This study - Wide-angle model with RSD} \\
    \hline\hline
     Field $ab$ & Term $n$ & $\Ai$ & $\Bi$ & $\Pabi$ \\
    \hline
    & $0$ & $(b\sigma_8) ^2$  & $\exp\left[-\frac{k^2 \sigma_{\mathrm{g}}^2 (\mu_1^2 + \mu_2^2)}{2}\right]$ & $P_{\mathrm{mm}}(k)$ \\
    $gg$ & $1$ & $ (b\sigma_8)^2 \beta_f$  & $(\mu_1^2 + \mu_2^2) \exp\left[-\frac{k^2 \sigma_{\mathrm{g}}^2 (\mu_1^2 + \mu_2^2)}{2}\right]$ & $P_{\mathrm{m\theta}}(k)$ \\
    & $2$ & $(b\sigma_8)^2 \beta_f^2$  & $\mu_1^2 \mu_2^2 \exp\left[-\frac{k^2 \sigma_{\mathrm{g}}^2 (\mu_1^2 + \mu_2^2)}{2}\right]$ & $P_{\mathrm{\theta\theta}}(k)$ \\
    \hline
    $gv$ & $0$ & $(b\sigma_8)^2 \beta_f$  & $(iaH) \frac{\mu_2}{k} \exp\left[-\frac{(k \sigma_{\mathrm{g}}\mu_1)^2}{2}\right]$ & $P_{\mathrm{m\theta}} (k) D_{\mathrm{u}}(k,\sigma_{\mathrm{u}})$ \\
    & $1$ & $(f\sigma_8)^2$  & $(iaH) \frac{\mu_2 \mu_1^2}{k} \exp\left[-\frac{(k \sigma_{\mathrm{g}}\mu_1)^2}{2}\right]$ & $P_{\mathrm{\theta\theta}} (k) D_{\mathrm{u}}(k,\sigma_{\mathrm{u}})$ \\
    \hline
    $vv$ & $0$ & $(f\sigma_8)^2$  & $(aH)^{2} \frac{\mu_1 \mu_2}{k^2}$ & $P_{\mathrm{\theta\theta}} (k) D_{\mathrm{u}}^2(k,\sigma_{\mathrm{u}})$ \\
    \hline\hline
    \end{tabular}}
    \tablefoot{The decomposition was performed in the \texttt{flip} framework following Eq.~\ref{eq:Pwflip}: The $\Ai$ terms correspond to the model parameters to fit, $\Bi$ to the geometrical terms that are analytically integrated, and $\Pi$ to the individual power spectrum terms that are numerically integrated. For the case of~\cite{carreres_growth-rate_2023} model which included only velocity covariance, the $vv$ terms corresponds to the one of \textbf{AB17}, \textbf{L22}, and \textbf{RC25}. In addition to all previously defined terms, we define the galaxy bias $b$, the RSD parameter $\beta_f = f/b$, the RSD FoG parameter $\sigma_{\mathrm{g}}$, and the velocity position FoG parameter, $\sigma_{\mathrm{u}}$. For the \textbf{L22}, we grouped the $gg$ terms with index $p$ and $q$ which have the same sum $m = p+q$. The $m$ index is run for each term depending on the maximal term in the Taylor expansion (i.e., $p$ and $q$ for $gg$, and $m$ for $gv$).}
\end{table*}

\endgroup

\FloatBarrier

\section{Numerical considerations}
\label{appendix:numerical}

\begin{figure}
	\includegraphics[width=0.95\columnwidth]{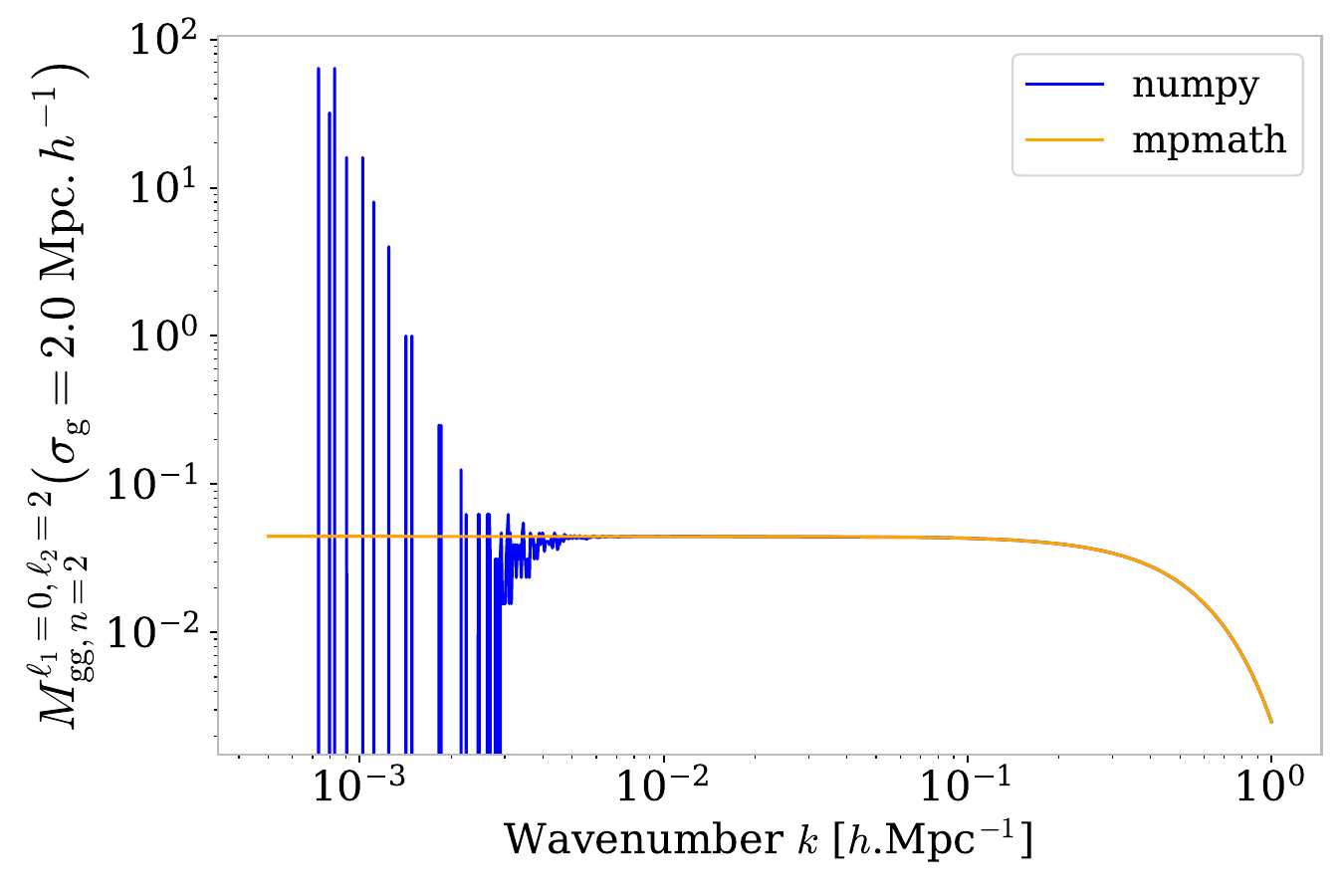}
    \caption{Wavenumber term $M_{\mathrm{gg},n=2}^{\ell_1=0, \ell_2=2}$(k) that is used in the computation of our covariance model \textbf{RC25}. The blue line is highly unstable at small wavenumber due to catastrophic cancellation with \texttt{numpy}. The orange curve shows the arbitrary precision calculation we are performing with \texttt{mpmath} Python package.}
    \label{fig:mll_numerical}
\end{figure}

For some models in table \ref{tab:models}, we need special care when computing the $\Mll(k)$ (or $\Mlnp(k)$ for plane-parallel) terms that are used in the computation of the field covariance. In particular, when an exponential term is included in the $\Bi$ part of the power spectrum model, for example for \textbf{AB20} or \textbf{RC25}, $\Mlnp(k)$ and $\Mll(k)$ terms correspond, respectively, to a linear combination of very large values when the product $k \sigma_{\mathrm{g}}$ reaches low values. For the \textbf{AB20} model, in Appendix A of \cite{adams_joint_2020}, the $K_{gg,n,l}$ terms (corresponding to the $M_{\mathrm{gg},n}^{\ell} N_{\mathrm{gg},\ell}$ product in the \texttt{flip} formalism) have terms that are proportional to $(k\sigma_{\mathrm{g}})^{-9}$. For the wide-angle model \textbf{RC25}, it can be proportional to $(k\sigma_{\mathrm{g}})^{-10}$ in the worst case. To illustrate this, we use the expression of $\Mll$ in our model \textbf{RC25} for the third  term ($n=2$) of the density-density ($gg$) correlation, for the $l_1 = 0$ and $l_2 =2$ case: 

\begin{align}
\begin{split}
    M_{\mathrm{gg},n=2}^{\ell_1=0, \ell_2=2} &= \frac{e^{- k^{2} \sigma_{\mathrm{g}}^{2}}}{k^{4} \sigma_{\mathrm{g}}^{4}} - \frac{\sqrt{2} \sqrt{\pi} e^{- \frac{k^{2} \sigma_{\mathrm{g}}^{2}}{2}} \operatorname{erf}{\left(\frac{\sqrt{2} k \sigma_{\mathrm{g}}}{2} \right)}}{4 k^{5} \sigma_{\mathrm{g}}^{5}}  \\
    &- \frac{\pi \operatorname{erf}^{2}{\left(\frac{\sqrt{2} k \sigma_{\mathrm{g}}}{2} \right)}}{4 k^{6} \sigma_{\mathrm{g}}^{6}} + \frac{9 e^{- k^{2} \sigma_{\mathrm{g}}^{2}}}{2 k^{6} \sigma_{\mathrm{g}}^{6}} -  \\
    &\frac{9 \sqrt{2} \sqrt{\pi} e^{- \frac{k^{2} \sigma_{\mathrm{g}}^{2}}{2}} \operatorname{erf}{\left(\frac{\sqrt{2} k \sigma_{\mathrm{g}}}{2} \right)}}{2 k^{7} \sigma_{\mathrm{g}}^{7}}+ \frac{9 \pi \operatorname{erf}^{2}{\left(\frac{\sqrt{2} k \sigma_{\mathrm{g}}}{2} \right)}}{4 k^{8} \sigma_{\mathrm{g}}^{8}}
\end{split}
\end{align}

In figure~\ref{fig:mll_numerical}, we show the shape of this function for the value $\sigma_{\mathrm{g}} = 2.0\ \mpc$. At low wavenumber, this function is a linear combination of very large floats (up to $10^{20}$) that cancel out to give a small value of the order of unity or less. It leads to what is sometimes called catastrophic cancellation, i.e., the small difference will be largely misestimated by \texttt{numpy} \git{numpy/numpy}{}~\citep{harris2020array}, leading to large numerical instabilities. To solve this issue, we use the \texttt{mpmath} \git{mpmath/mpmath}{}~\citep{mpmath} Python package, which allows us to make floating-point arithmetic calculations with arbitrary precision. We use this package for all the $\Mll$, which integrates exponential RSD FoG contaminants terms, as shown in the figure~\ref{fig:mll_numerical} for the aforementioned term. We check that the \texttt{mpmath} solution is enough to cover all reasonable values of $\sigma_{\mathrm{g}}$ (typically higher than 1 $\mpc$) without instabilities.

A second numerical issue encountered by our formalism is the instabilities introduced by the use of Hankel transforms in the numerical integration of the equation~\ref{eq:cov_flip_wa} (or~\ref{eq:cov_flip_pp} in the plane-parallel case). When performing a Hankel transform, the wavenumber range should cover all the separations ($r = |\mathbf{r}|$) inside the covariance matrix element. For the case of small separations ($k \sim 1.0\ h\cdot\mathrm{Mpc}^{-1}$), we choose to switch to standard Simpson's numerical integration, which is costly but only for a small number of terms in the full covariance. Note that the linear modeling theories we are using are insufficient for those small separations, and we plan to improve nonlinear corrections in future works.

For the special case when we are using a simulation, if the minimal wavenumber of the power spectrum $\Pabi$ employed to perform covariance calculation is lower than the maximal size of the box, the Hankel transform accounts for a too large number of modes in the integration. It adds additional terms to the field covariance matrix which can make it non-positive defined. We solved this issue by computing the numerical integration with standard Simpson's numerical integration between the lowest wavenumber of the power spectrum and a wavenumber $k_{\mathrm{min}}$, corresponding to the lowest mode of the considered simulation. We then subtract this additional integral from the Hankel transform to remove those terms. This regularization should only be used for simulations, not for data containing the Universe's large-scale wavenumber modes. 

\section{Wide-angle impact on covariance contraction}
\label{appendix:waimpact}

\begin{figure*}
	\includegraphics[width=0.95\textwidth]{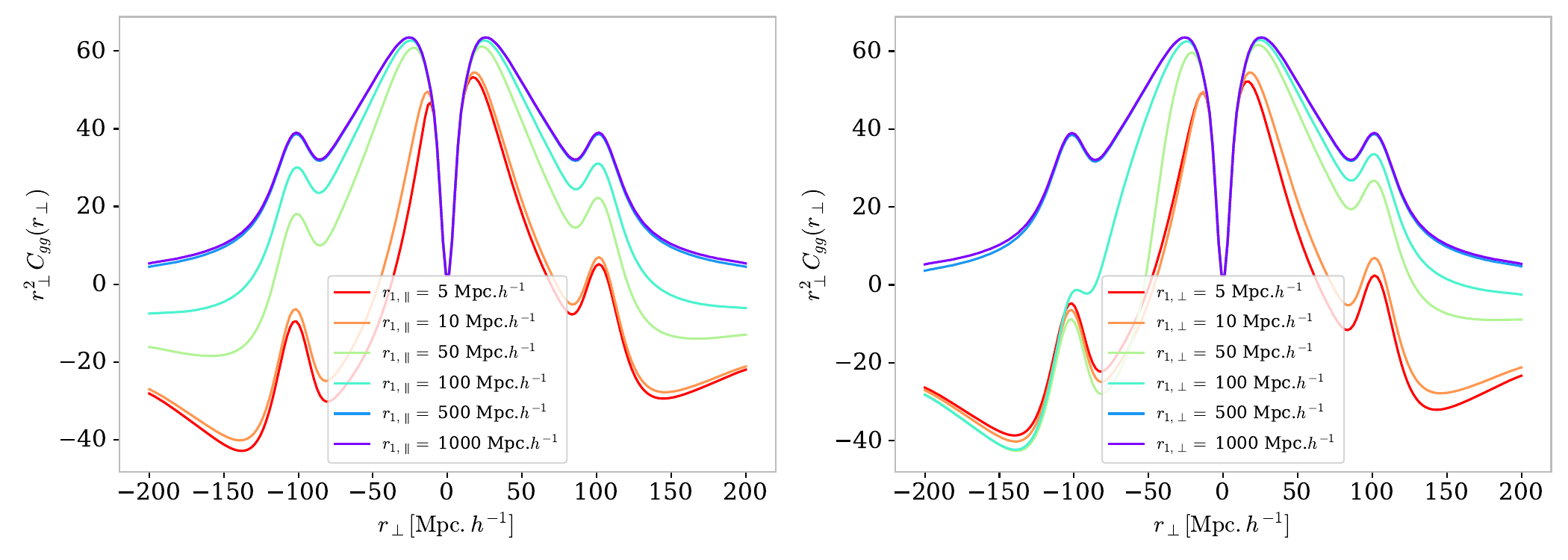}
    \caption{Galaxy-galaxy covariance contraction for the \textbf{RC25} model with the same parameterization than \ref{fig:contraction} (and $\sigma_{\mathrm{g}} = 3.0\ \mpc$) while varying the reference position $\mathbf{r_1}$ in the parallel $r_{1,\parallel}$ (left) and perpendicular $r_{1,\bot}$ (right) direction. Those contractions are expressed in one-dimension as a function of perpendicular separation $r_{\bot}$ when $r_{\parallel} = 0\ \mpc$.}
    \label{fig:wide_angle_impact}
\end{figure*}

\begin{figure*}
	\includegraphics[width=0.95\textwidth]{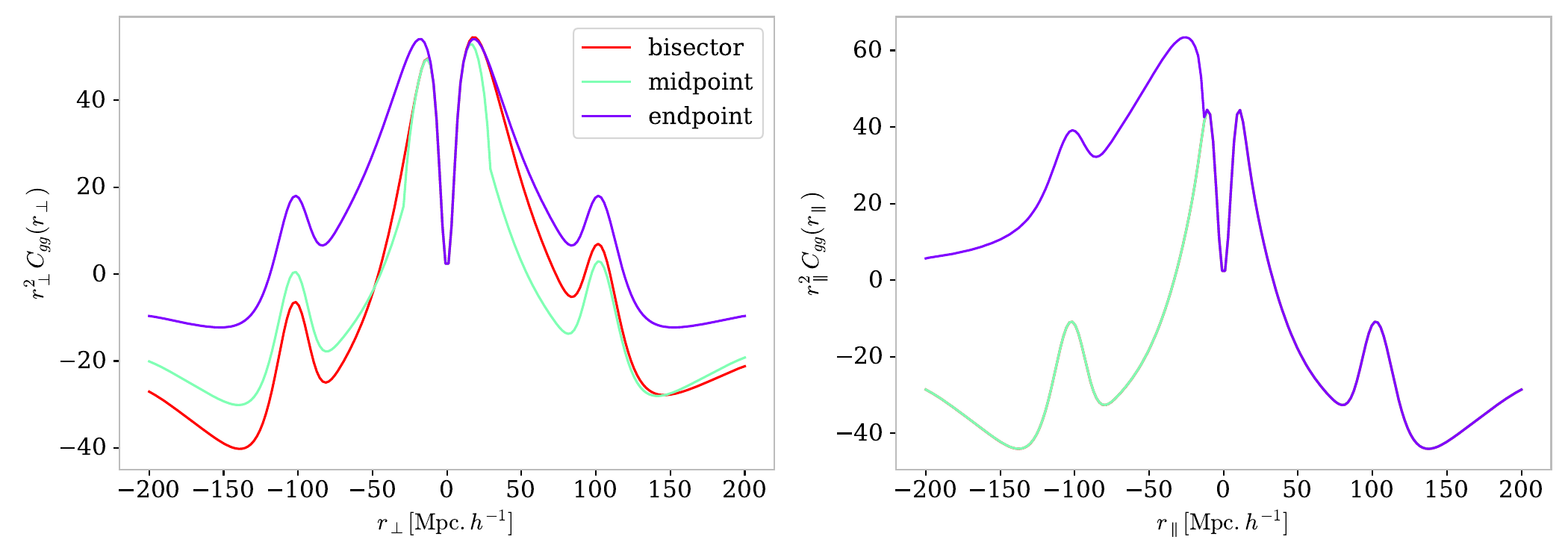}
    \caption{Galaxy-galaxy covariance contraction for the \textbf{RC25} model with the same parameterization than \ref{fig:wide_angle_impact} while varying the reference position the definition for the line-of-sight reference vector $\mathbf{d}$ between bisector, midpoint, and endpoint. The reference vector taken is $r_{1,\bot} = 10\ \mpc$~and $r_{1,\parallel} = 10\ \mpc$.}
    \label{fig:d_definition}
\end{figure*}

This section shows complementary tests conducted on the impact of wide-angle on covariance matrix calculation.

Figure~\ref{fig:wide_angle_impact} gives the effect of varying the reference position $\mathbf{r_1}$ on the $gg$ covariance contraction computed with the \textbf{RC25} model. This impact is only shown as a function of perpendicular separation $r_{\bot}$ because the wide-angle effects in the covariance calculation is not sensitive to parallel separation when the perpendicular separation is null. Both for large $r_{1,\bot}$ and $r_{1,\parallel}$ values, the contraction gives the shape of the plane-parallel limit shown in figure~\ref{fig:contraction}. At low values of reference position, the impact of wide-angle is significant and gives a non-symmetric shape with respect to the separation.

Figure~\ref{fig:d_definition} shows the same covariance contraction but varying the definition of the line-of-sight reference vector $\mathbf{d}$ in equation~\ref{eq:d_def}. It is computed for $r_{1,\bot} = 10\ \mpc$~and $r_{1,\parallel} = 10\ \mpc$ in order to obtain large wide-angle effects. The endpoint definition is unsuitable for $\mathbf{d}$ as it gives non-symmetric contraction as a function of parallel separation $r_{\parallel}$ for which the covariance should not be sensitive. Furthermore, it does not exhibit the expected asymmetric behavior over the perpendicular separation axis. Endpoint is the less-used definition for wide-angle correction, as it requires arbitrarily selecting one point as a reference over the other. The midpoint and bisector definition gives a similar contraction profile, but the midpoint definition shows a slightly jagged line pattern near $25\ \mpc$ separation. This test convinces us to adopt the bisector definition, which is already the most widely used for wide-angle correlation computation.

\end{appendix}

\end{document}